\documentclass[journal]{IEEEtran}
\usepackage{cite}
\usepackage{amsmath,amssymb,amsfonts}
\usepackage{algorithmic}
\usepackage{algorithm}
\usepackage{graphicx}
\usepackage{stmaryrd}
\usepackage{xcolor}
\usepackage{array}
\usepackage{commath}
\usepackage{sidecap}
\usepackage{stfloats}
\usepackage{tabularx, boldline}
\usepackage{rotating,booktabs,multirow}
\usepackage{mathtools}
\usepackage{flexisym}
\usepackage{breqn}
\usepackage{url}
\usepackage{adjustbox}
\usepackage{makecell}

\usepackage{textcomp}
\usepackage{amsmath}
\usepackage{txfonts}
\usepackage{graphicx}%,subcaption}
\usepackage{acronym}
\usepackage{tabularx, boldline}
\usepackage{rotating,booktabs,multirow}
\usepackage{enumerate}
\usepackage{etoolbox}
\usepackage[]{nomencl}   
    \makenomenclature

\providetoggle{nomsort}
\settoggle{nomsort}{true} % true = sort by use, false = sort as usual

\makeatletter
\iftoggle{nomsort}{%
    \let\old@@@nomenclature=\@@@nomenclature        
        \newcounter{@nomcount} \setcounter{@nomcount}{0}%
        \renewcommand\the@nomcount{\two@digits{\value{@nomcount}}}% Ensure 10>01
        \def\@@@nomenclature[#1]#2#3{% Taken from package documentation
          \addtocounter{@nomcount}{1}%
        \def\@tempa{#2}\def\@tempb{#3}%
          \protected@write\@nomenclaturefile{}%
          {\string\nomenclatureentry{\the@nomcount\nom@verb\@tempa @[{\nom@verb\@tempa}]%
          \begingroup\nom@verb\@tempb\protect\nomeqref{\theequation}%
          |nompageref}{\thepage}}%
          \endgroup
          \@esphack}%
      }{}
\makeatother

\def\R{\mathbb{R}} % Números reales
\def\E{\mathbb{E}} % Esperanza
\newcommand{\mkv}{-\!\!\!\!\minuso\!\!\!\!-}

\usepackage{authblk}
\def\BibTeX{{\rm B\kern-.05em{\sc i\kern-.025em b}\kern-.08em
    T\kern-.1667em\lower.7ex\hbox{E}\kern-.125emX}}

\ifCLASSINFOpdf

\else

\fi

\begin{document}

%\title{Optimal Privacy-Utility Trade-off with Directed Information as Privacy Measure}
%\title{A Deep Recurrent Adversarial Learning Approach for Privacy-Preserving Smart Meter Data Release}

%\title{Deep Recurrent Adversarial Learning for Privacy-Preserving Smart Meter Data Release}
\title{Real-Time Privacy-Preserving Data Release\\ for Smart Meters}

\author{Mohammadhadi~Shateri,~\IEEEmembership{Member,~IEEE,} Francisco~Messina,~\IEEEmembership{Member,~IEEE,}
        Pablo~Piantanida,~\IEEEmembership{Senior Member,~IEEE,}
        Fabrice~Labeau,~\IEEEmembership{Senior Member,~IEEE}%
%\author{Mohammadhadi~Shateri,~\IEEEmembership{Member,~IEEE,}, Francisco~Messina,~\IEEEmembership{Member,~IEEE,}
%        Pablo~Piantanida,~\IEEEmembership{Senior Member,~IEEE,}
%        Fabrice~Labeau,~\IEEEmembership{Senior Member,~IEEE}% <-this % stops a space
%\thanks{This work was supported by Hydro-Quebec, the Natural Sciences and Engineering Research Council of Canada, and McGill University in the framework of the NSERC/Hydro-Quebec Industrial Research Chair in Interactive Information Infrastructure for the Power Grid (IRCPJ406021-14). The work of Prof. Pablo Piantanida was supported by the European Commission’s Marie Sklodowska-Curie Actions (MSCA), through the Marie Sklodowska-Curie IF (H2020-MSCAIF-2017-EF-797805-STRUDEL). We would like to acknowledge support for this project from the CNRS via the International Associated Laboratory (LIA) on \emph{Information, Learning and Control}. }
\thanks{M. Shateri, F. Messina, and F. Labeau  are with the Department of Electrical and Computer Engineering, McGill University, QC, Canada.\\ Email:\{mohammadhadi.shateri,francisco.messina\}@mail.mcgill.ca}% <-this % stops a space
\thanks{P. Piantanida is with Laboratoire des Signaux et Syst\`emes, CentraleSup\'elec-CNRS-Universit\'e Paris Sud, Gif-sur-Yvette, France and with Montreal Institute for Learning Algorithms (Mila), Universit\'e de Montr\'eal, QC, Canada.}% <-this % 
}

%{Shell \MakeLowercase{\textit{et al.}}: Bare Demo of IEEEtran.cls for IEEE Transactions on Magnetics Journals}
\markboth{IEEE Transactions on Smart Grid}{}

\maketitle


\begin{abstract}

Smart Meters (SMs) are able to share the power consumption of users with utility providers almost in real-time. These fine-grained signals carry sensitive information about users, which has raised serious concerns from the privacy viewpoint. In this paper, we focus on real-time privacy threats, i.e., potential attackers that try to infer sensitive information from SMs data in an online fashion. We adopt an information-theoretic privacy measure and show that it effectively limits the performance of any attacker. Then, we propose a general formulation to design a privatization mechanism that can provide a target level of privacy by adding a minimal amount of distortion to the SMs measurements. On the other hand, to cope with different applications, a flexible distortion measure is considered. This formulation leads to a general loss function, which is optimized using a deep learning adversarial framework, where two neural networks --referred to as the releaser and the adversary-- are trained with opposite goals. An exhaustive empirical study is then performed to validate the performance of the proposed approach and compare it with state-of-the-art methods for the occupancy detection privacy problem. Finally, we also investigate the impact of data mismatch between the releaser and the attacker.% the .} %Finally, the impact of data mismatch (between the model and attacker) in degrading the performance of an attacker is studied.}

\end{abstract}


% \fi
\IEEEpeerreviewmaketitle

\nomenclature{SMs}{Smart Meters}
\nomenclature{MI}{Mutual Information}
\nomenclature{DI}{Directed Information}
\nomenclature{RNNs}{Recurrent Neural Networks}
\nomenclature{LSTM}{Long-Short Term Memory}
\nomenclature{i.i.d.}{Independent and Identically distributed} 
\nomenclature{$X^T = (X_1, \ldots, X_T)$}{A sequence of random variables, or a time series, of length $T$}
\nomenclature{$x^T = (x_1, x_2, \ldots, x_T)$}{A realization of $X^T$}
\nomenclature{$x^{(i)T} = (x^{(i)}_1, x^{(i)}_2, \ldots, x^{(i)}_T)$}{ $i^{\text{th}}$ sample in a minibatch used for training}
\nomenclature{$\E[X]$}{ Expectation of a random variable $X$}
\nomenclature{$p_X$}{Probability distribution of $X$}
\nomenclature{$I(X;Y)$}{Mutual information between $X$ and $Y$}
\nomenclature{$H(X)$}{Entropy of random variable $X$}
\nomenclature{$I(X^T \to Y^T)$}{Directed information between $X^T$ and $Y^T$}
\nomenclature{${X \mkv Y \mkv Z}$}{Markov chain among $X$, $Y$ and $Z$}
\nomenclature{$\inf(.)$}{Infimum}

\printnomenclature

\section{Introduction}\label{intro}

\subsection{Motivation}

\IEEEPARstart{S}{M}s are a cornerstone for the development of smart electrical grids. These devices are able to report power consumption measurements of a house to a utility provider every hour or even every few minutes. This feature generates a considerably amount of useful data which enables several applications in almost real-time such as power quality monitoring, timely fault detection, demand response, energy theft prevention, etc.~\cite{alahakoon2016,wang2019}. However, this fine-grained power consumption monitoring poses a threat to consumers privacy. As a matter of fact, it has been shown that simple algorithms, known in general as NonIntrusive Load Monitoring (NILM) methods, can readily be used to infer the types of appliances being used at a household at a given time from the SMs data~\cite{molina2010}. Since these features are highly correlated with the presence of people at the dwelling and their personal habits, this induces serious privacy concerns which can have an impact on the acceptance and deployment pace of SMs~\cite{mckenna2012}. The natural challenge raised here is: how can privacy be enhanced while preserving the utility of the SMs data? Although the privacy problem has been widely studied in the field of data science~\cite{jain2016}, the time series structure of SMs data requires a particular treatment~\cite{asghar2017}. For a recent survey about SMs privacy, the reader is \mbox{referred to~\cite{giaconi2018privacy}}.

\subsection{Related work}

Simple approaches for preservation of privacy in the context of SMs include data aggregation and encryption~\cite{li2010,rottondi2013}, the use of pseudonyms rather than the real identities of users~\cite{efthymiou2010smart}, downsampling of the data~\cite{mashima2015authenticated,eibl2015} and random noise  addition~\cite{barbosa2016}. However, these methods often restrict the potential applications of the SMs data in an uncontrolled way. For instance, downsampling of the data may incur time delays to detect critical events, while data aggregation degrades the positioning and accuracy of the power measurements.

A formal approach to the privacy problem has been presented in~\cite{sankar2013smart} from an information-theoretic perspective, where it has been proposed to assess privacy by the MI between the sensitive variables to be hidden and the power measurements distorted by a privatizer mechanism. More specifically, the authors model the power measurements of SMs with a hidden Markov model in which the distribution of the measurements is controlled by the state of the appliances, and for each particular state, the distribution of power consumption is assumed to be Gaussian. This model is then used to obtain the privacy-utility trade-off using tools from rate-distortion theory~\cite{cover2006elements}. Although this approach is very appealing, it has two important limitations for its application to real-time privacy problems using actual data. First, the privacy-preserving data release mechanism works with blocks of samples, which is not well-suited for real-time processing. Second, the Gaussian model may be quite restrictive to model SMs signals. The information-theoretic approach was used in other privacy-aware SMs studies such as~\cite{zheng2017privacy} where the MI between the distorted SMs data and sensitive appliance states at time slot $t$ was considered as the privacy measure. However, in this work, the temporal correlation in SMs data is not taken into account.

More sophisticated approaches consider the use of Rechargeable Batteries (RBs) and Renewable Energy Sources (RES) in homes in order to modify the actual energy consumption of users with the goal of hiding the sensitive information~\cite{kalogridis2010,tan2013,zhang2016cost,sun2017smart,li2018information,giaconi2018}. The main motivation to introduce the use of physical resources into the privacy problem comes from the observation that this strategy does not require any distortion in the actual SMs measurements, which means that there is no loss in terms of utility. However, the incorporation of physical resources may not only make the problem more complex and limited in scope, but can also generate a significant cost to users due to the faster wear and tear of the RBs as a consequence of the increased charging/discharging rate~\cite{giaconi2018privacy}. On the other hand, the required level of distortion for a specific privacy goal in a realistic scenario in which the attacker threatening privacy has only partial information is still an open question. Thus, the need and convenience of these solutions is still questionable. However, it is also important to note that these approaches are complementary to the ones based on distorting the power measurements.

The use of neural networks to model a privacy attacker has been considered in~\cite{wang2017}. However, a more powerful formulation of the problem is obtained if one assumes that both the releaser (i.e., the privatizer) and the attacker are deep neural networks (DNNs). In this framework, the releaser can be trained by simulating an attacker based on a minimax game, an idea that is inspired by the well-known Generative Adversarial Networks (GANs)~\cite{goodfellow2014}. This concept can be referred to as Generative Adversarial Privacy (GAP)~\cite{huang2018} and is the basis for the approach taken in the present work. It should be mentioned that the concept of GAP has been studied for different applications related to image classification~\cite{tripathy2019privacy,feutry2018} but, to the best of our knowledge, not in the context of SMs. In these works, the authors consider independent and identically distributed (i.i.d.) data and deep feed-forward neural networks for the releaser and attacker, which are unable to capture and exploit the time correlation in the time series SMs signals. The idea of time-series generation with an adversarial approach has been considered in~\cite{esteban2018} for medical data based in the principle of differential privacy.

\subsection{Contributions}

In this paper, we adopt a distortion-based real-time privacy-preserving strategy. For simplicity, we assume that no RBs and/or RESs are available and thus, the distortion on power measurements is the only mean to achieve a desired privacy level. The main contributions of this work, which is an extension of a short version in~\cite{shateri2019deep}, are the following:

\begin{enumerate}[(i)]	

	\item We applied DI as a privacy measure and show its theoretical relevance for the privacy problem under consideration. It is worth to mention that DI was first used in~\cite{erdogdu2015} but in a different manner. In addition, unlike this and other works such as~\cite{sankar2013smart} , we impose no explicit assumptions on the generating model of the power measurements, but take a more versatile data-driven approach.

	\item We study different distortion measures to provide more flexibility to control the specific features to be preserved in the released signals, i.e., the relevant characteristics for the targeted applications of the data.
	
	\item  For the sake of computational tractability, we propose a loss function for training the privacy-preserving releaser based on an upper bound of the DI. Then, considering an attacker that minimizes a Kullback-Leibler divergence between the true and approximate distributions of the sensitive variables given the released signal, we provide a relaxed formulation of the original problem suitable for a deep learning framework.
	
	\item We perform an extensive statistical study with actual data to characterize the utility-privacy trade-offs and the nature of the distortion generated by the releaser network.
	 
	\item We investigate the data mismatch problem in the context of SMs privacy, which occurs when the data available to the attacker is not the same as the one used for training the releaser mechanism, and show that it has an important impact on the privacy-utility trade-off.	This confirms that, under some conditions, the privacy-utility trade-off can indeed be much less severe than expected.
	
	\item To the best of our knowledge, this is the first time that the concept of the generative adversarial privacy is used in the context of the SMs data privacy preservation. In addition, in this paper we consider deep RNNs to capture and exploit the time correlation of SMs signals.
	
\end{enumerate}

\subsection{Organization of the paper}

The rest of the paper is organized as follows. In Section \ref{sec:formulation}, we present the theoretical formulation of the problem. This leads to the loss functions for the releaser and attacker posed in Section \ref{sec:model}, where the privacy-preserving adversarial framework is introduced along with the training algorithm. Extensive results are presented and discussed in Section \ref{sec:results}. Finally, some concluding remarks are presented in Section \ref{sec:conclusion}.

\section{Problem Formulation}
\label{sec:formulation}

\begin{figure}[htbp!]
	\centering
	\includegraphics[width=0.97\linewidth]{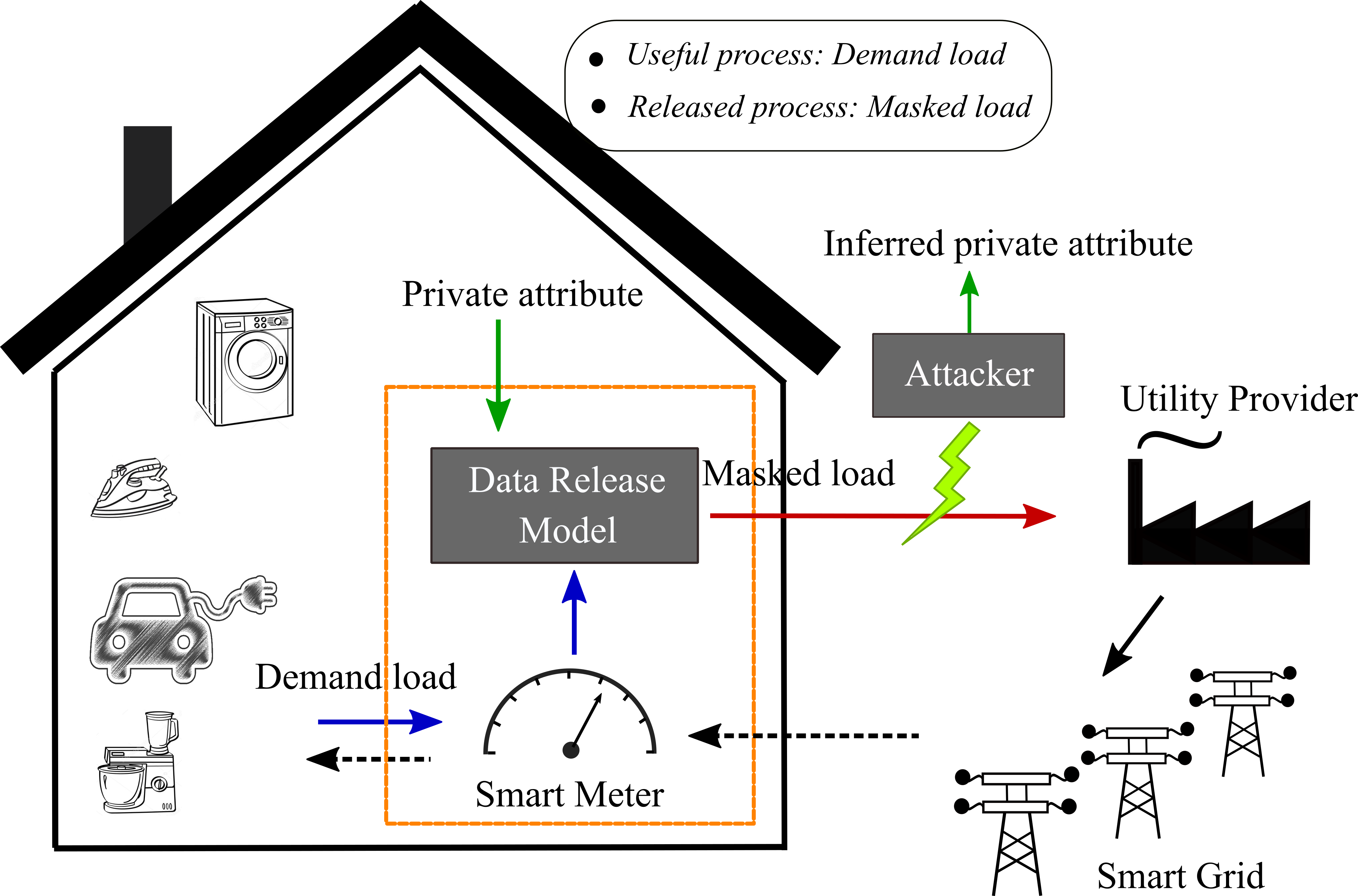}
	\caption{Privacy-preserving data release framework for SMs.}
	\label{fig:SMs-privacy-setting}
\end{figure}

The privacy-preserving framework studied in this paper is presented in  Fig.\ref{fig:SMs-privacy-setting}. As shown, a data releaser (which is aware of the private attributes that the user wants to hide) manipulates the actual SM measurements before sharing them to the UP to prevent leakage of sensitive information that could be inferred by a malicious attacker such as an eavesdropper.

There are four main types of variables that should be clearly defined in the privacy-preserving data release setting: (i) the private/sensitive attribute which we aim to hide $X^T$ (e.g., occupancy state of a house over time); (ii) the useful process for the utility provider $Y^T$ (e.g., actual electricity consumption of the household), which is generally highly correlated with the private data; (iii) the observed signal $W^T$, a combination of private and useful variables, which is the input to the data release system; (iv) and the released process $Z^T$, a sanitized version of $Y^T$ which is the output of the data release system. 

We assume that $X_t$ takes values on a fixed discrete alphabet $\mathcal{X}$ for each $t \in \{ 1, \ldots, T \}$. At each time $t$, a releaser produces the released process $Z_t$ based on the observation $W^t$, while an attacker attempts to infer $X_t$ based on $Z^t$ by using an approximation of $p_{X^T|Z^T}$, which we shall denote by $p_{\hat{X}^T|Z^T}$. Notice that the releaser must be causal in order to avoid delays in the SMs data report process. In addition, we treat the case in which the attacker is performing the inference in real-time, so that it is also causal. This assumption is reasonable for scenarios in which the sensitive information is valuable in a timely manner (e.g., in the case of targeted burglary based on occupancy detection \cite{asghar2017}). However, it should be noted that not all privacy threats fall under this umbrella. Alternative attacker structures, which may be interesting in other scenarios, are left for future work and are out of the scope of this paper. Note that due to the previous assumption the distribution $p_{Z^T\hat{X}^T|W^T}$ can be decomposed as follows:

\begin{equation} \label{eq:distribution_structure} p_{Z^T\hat{X}^T|W^T}(z^T,\hat{x}^T|w^T) = \prod_{t=1}^{T} p_{Z_t|W^t}(z_t|w^t) p_{\hat{X}_t|Z^t}(\hat{x}_t|z^t). \end{equation}

In abstract terms, the goal of the releaser is to minimize the information leakage of the sensitive process $X^T$ while simultaneously keeping the distortion between the released time series $Z^T$ and the useful signal $Y^T$ small. On the other hand, the goal of the attacker is to infer $p_{X_t|Z^t}$, for each $t$, as accurately as possible. Note that after the approximation $p_{\hat{X}_t|Z^t}$ is obtained, the attacker can estimate the private information $x^t$ from observations $z^t$ in an online (causal) fashion, by solving %the following $T$ problems:
\begin{equation} 
\underset{\hat{x}_t \in \mathcal{X}}{\text{argmax}} \; p_{\hat{X}_t|Z^t}(\hat{x}_t|z^t), 
\end{equation} % \qquad t \in \{1,\ldots,T\}. 
at each $t=1,\dots,T$. Thus, the attacker can be interpreted as a classifier or hypothesis test, as stated in \cite{li2019}. However, in the present case, we consider the more realistic and general scenario in which the statistical test is sub-optimal due to the fact that the attacker has no access to the actual conditional distributions $p_{X_t|Z^t}$ but only to its approximation $p_{\hat{X}_t|Z^t}$.

In order to take into account the causal relation between $X^T$ and $\hat{X}^T$, the information leakage is quantified by the DI~\cite{massey1990causality}: %quantities 
\begin{equation} \label{eqtr1} I\big(X^T\rightarrow \hat{X}^T\big)\coloneqq \sum_{t=1}^{T} I(X^t;\hat{X}_t|\hat{X}^{t-1}), \end{equation}
where $I(X^t;\hat{X}_t|\hat{X}^{t-1})$ is the conditional MI between $X^t$ and $\hat{X}_t$ conditioned on $\hat{X}^{t-1}$~\cite{cover2006elements}. 

The normalized expected distortion between $Y^T$ and its noisy (or disturbed) observation $Z^T$ is defined as:
\begin{equation} \label{Distortion} D(Z^T,Y^T) \coloneqq \frac{\E[d(Z^T,Y^T)]}{T}, \end{equation}
where $d : \R^T \times \R^T \to \R$ is any distortion function (i.e., a metric on $\R^T$). To ensure the quality of the release, it is natural to impose the following constraint: $D(Z^T,Y^T) \le \varepsilon$ for some given $\varepsilon \ge 0$. In previous works, the normalized squared-error was considered as a distortion function (e.g., \cite{sankar2013smart,shateri2019deep}). Nevertheless, other distortion measures can also be relevant within the framework of SMs. For instance, demand response programs usually require an accurate knowledge of peak power consumption, so a distortion function closer to the infinity norm would be more meaningful for those particular applications. Thus, for the sake of generality and to keep the distortion function simple, we propose to use an $\ell_p$ distance:
\begin{equation} \label{eq:p_distortion} d(z^T,y^T) \coloneqq \| z^T - y^T \|_p = \left( \sum_{t=1}^T |z_t - y_t|^p \right)^{1/p}, \end{equation}
where $p \ge 2$ is a fixed parameter. Note that this distortion function leads to the root-mean-squared error when $p = 2$, while it converges to the maximum error between the components of $z^T$ and $y^T$ as $p \to \infty$.

Therefore, the problem of finding an optimal releaser subject to the optimal (Bayesian) attacker and distortion constraint can be formally written as follows:
\begin{align} \label{eqtr2} \underset{p_{Z^T|W^T}}{\text{inf}}  \;  I\left(X^T\rightarrow \hat{X}^T\right) \quad \text{subject to} \quad D(Z^T,Y^T) \le \varepsilon. \end{align}
Note that the solution to this  optimization problem requires a balance between the attacker $p_{\hat{X}^T|Z^T}$ and the releaser $p_{Z^T|W^T}$, where the optimal attacker consists in  inferring the private information $\hat{X}^T\approx {X}^T$  and thus, the attacker  attempts to minimize the Kullback-Leibler divergence \cite{cover2006elements} between the  corresponding predictors:

\begin{align} \label{eq:attacker_optimization_problem} \underset{p_{\hat{X}^t|Z^t}}{\text{inf}} \;  \text{KL}\big (p_{X^t Z^t}\|p_{\hat{X}^t Z^t}\big ) & = \underset{p_{\hat{X}^t|Z^t}}{\text{inf}} \; \E\left[ \log \frac{p_{X^t|Z^t}(X^t|Z^t)}{p_{\hat{X}^t|Z^t}(X^t|Z^t)} \right], 
\end{align}
where the expectation is with respect to $p_{X^T Z^T}$. Note that solving \eqref{eq:attacker_optimization_problem} is equivalent to minimizing $\E[-\log p_{\hat{X}^t|Z^t}(X^t|Z^t)]$, the so-called cross-entropy.

\begin{figure}[htbp!]
	\centering
	\includegraphics[width=0.75\linewidth]{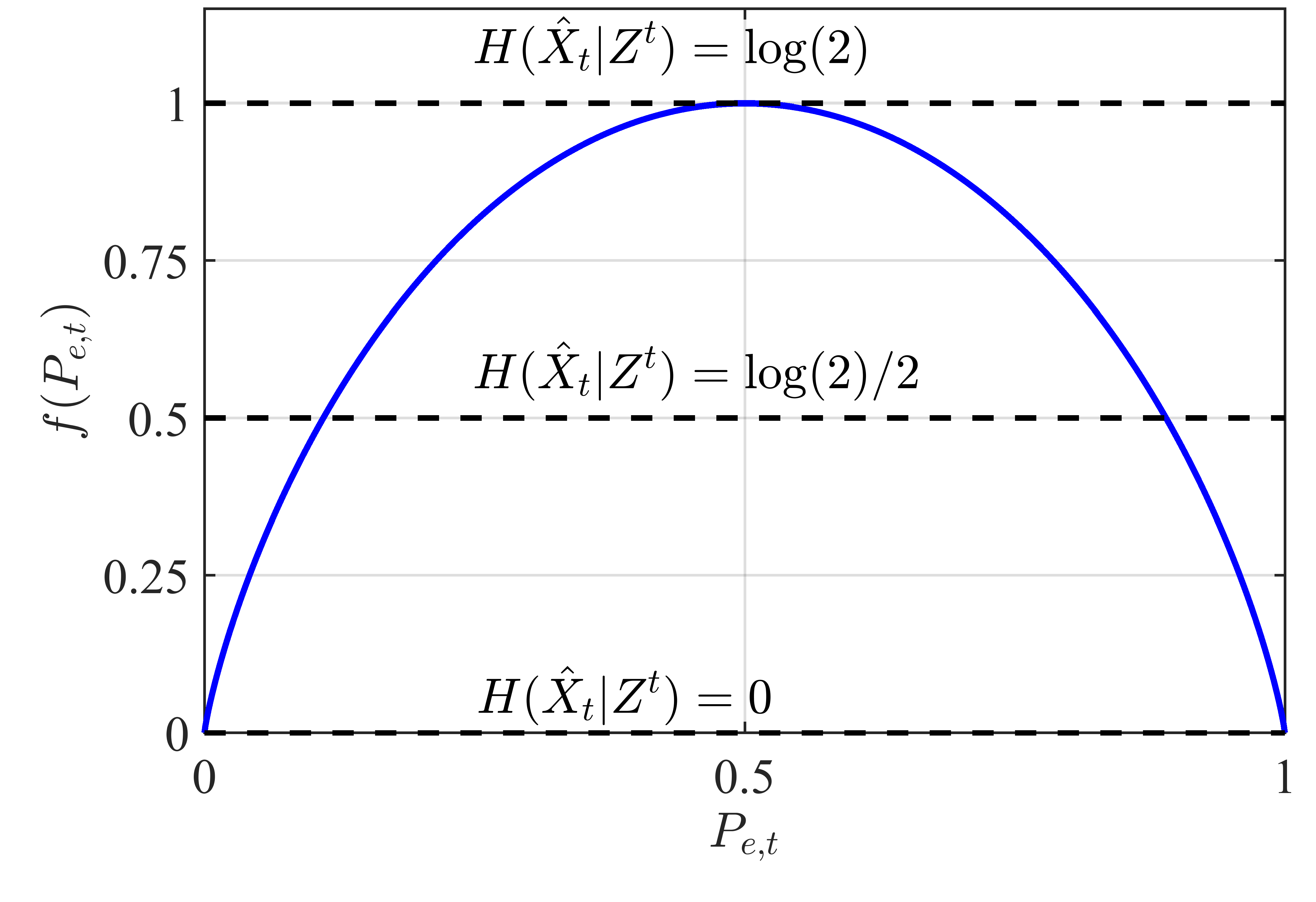}
	\caption{Plot of the Fano's function $f(P_{e,t})$ for the case $|\mathcal{X}| = 2$.}
	\label{fig:fano_bound}
\end{figure}

Unfortunately, the optimization problem \eqref{eqtr2} is, in general, computationally intractable. For instance, simply computing the DI would take $O(|\mathcal{X}|^T)$ operations, where $|\mathcal{X}|$ is the size of $\mathcal{X}$, which makes it not scalable for large sequences of data. However, it can be exploited to obtain a more convenient surrogate objective function for the releaser, by considering the following simpler upper bound:

\begin{align} %\label{eqtr5}
%\begin{aligned}
I\left(X^T\rightarrow \hat{X}^T\right) & = \sum_{t=1}^{T}\left[H(\hat{X}_t|\hat{X}^{t-1})-H(\hat{X}_t|\hat{X}^{t-1},X^t)\right]
\nonumber \\  &\overset{\text{(i)}}{\leq}\sum_{t=1}^{T}\left[H(\hat{X}_t|\hat{X}^{t-1})-H(\hat{X}_t|\hat{X}^{t-1},X^t,Z^t)\right] \nonumber \\ 
&\overset{\text{(ii)}}{=} \sum_{t=1}^{T} \left[ H(\hat{X}_t|\hat{X}^{t-1}) - H(\hat{X}_t|Z^t) \right]\nonumber \\ \label{eq:di_upper_bound}
& \overset{\text{(iii)}}{\leq} T \log|\mathcal{X}| - \sum_{t=1}^{T}H(\hat{X}_t|Z^t), 
\end{align}
where (i) is due to the fact that conditioning reduces entropy; equality (ii) is due to the Markov chain $(X^t,\hat{X}^{t-1}) \mkv Z^t \mkv \hat{X}_t$, which follows from \eqref{eq:distribution_structure}; and (iii) is due to the trivial bounds ${H(\hat{X}_t|\hat{X}^{t-1}) \le H(\hat{X}_t) \le \log(|\mathcal{X}|)}$. Note that minimizing the upper bound \eqref{eq:di_upper_bound} corresponds to maximizing $\sum_{t=1}^{T} H(\hat{X}_t|Z^t)$, which amounts to maximizing the total uncertainty of the attacker. In fact, from Fano's inequality \cite{cover2006elements}, we have that

\begin{equation} \label{eq:fano_bound} f(P_{e,t}) \coloneqq h(P_{e,t}) + P_{e,t} \log(|\mathcal{X}| - 1) \ge H(\hat{X}_t|X_t) \ge H(\hat{X}_t|Z^t), \end{equation}
where $P_{e,t} \coloneqq \mathbb{P}(X_t \ne \hat{X}_t)$ and $h(p) \coloneqq - p \log(p) - (1-p) \log(1-p)$ is the so-called binary entropy function. In addition, the bound is tight (i.e., it can not be strengthened without further assumptions) \cite{cover2006elements,ho2010}. It should be noted that this bound constrains $P_{e,t}$ to be in an interval around $P_{e,t} = (|\mathcal{X}| - 1)/|\mathcal{X}|$, which corresponds to the performance of an attacker that does uniform random guessing to infer the value of $X_t$. Indeed, in the extreme case in which $H(\hat{X}_t|Z^t) = \log(|\mathcal{X}|)$, we have that $P_{e,t} = (|\mathcal{X}| - 1)/|\mathcal{X}|$. Physically, this means that $Z^t$ is not providing any information to the attacker to infer $X_t$. Moreover, the length of this interval decreases monotonically when $H(\hat{X}_t|Z^t)$ is increased. Fig. \ref{fig:fano_bound} is presented to illustrate the proposed scenario for the binary case. From Fig. \ref{fig:fano_bound} it can be noticed that as $H(\hat{X}_t|Z^t)$ increases, the interval over which $P_{e,t}$ lies shrinks. Also, in the extreme case in which $H(\hat{X}_t|Z^t) = \log(2)$, Fano's inequality \eqref{eq:fano_bound} implies ${P_{e,t} = 1/2}$, which corresponds to the performance of random guessing the value of $X_t$, meaning that $Z^t$ does not provide any valuable information. Therefore, a releaser which attempts to maximize $H(\hat{X}_t|Z^t)$ is trying to constrain $P_{e,t}$ to be close to random guessing performance. This can be considered as a universal privacy guarantee and justifies the usefulness of the DI and, in particular, the surrogate upper bound \eqref{eq:di_upper_bound}.

Therefore, in this work, the information leakage is measured by the following average conditional entropy (ignoring constant terms):
 
\begin{equation} \label{eq:di_criterion} -\frac{1}{T} \sum_{t=1}^T H(\hat{X}_t|Z^t), \end{equation}
where the factor $1/T$ has been introduced for normalization purposes. It is interesting to notice that this is different from the formulation in \cite{sankar2013smart}, in which the authors consider the MI $I(X^T;Z^T) = H(X^T) - H(X^T|Z^T)$ as the information leakage measure, which can be equivalently written as follows (again ignoring constant terms and normalizing):

\begin{equation} \label{eq:mi_criterion} - \frac{1}{T} \sum_{t=1}^T H(X_t|Z^T,X^{t-1}). \end{equation}
By comparing \eqref{eq:di_criterion} and \eqref{eq:mi_criterion}, the differences between the two privacy measures are clear. The fact that we have assumed that the attacker has a causal structure explains why $Z^t$ appears in \eqref{eq:di_criterion} instead of $Z^T$ as in \eqref{eq:mi_criterion}. More fundamentally, the expression \eqref{eq:mi_criterion} corresponds to assuming that the attacker is optimal (i.e., $p_{\hat{X}_t|Z^t} = p_{X_t|Z^t}$) and has access to $X^{t-1}$ to infer $X_t$. These latter assumptions are not expected to hold in practice.

\section{Privacy-Preserving Model} \label{sec:model}

We now shift the focus from the abstract theoretical formulation to a practical one based on deep learning ideas. In particular, in this section, we model the releaser $p_{Z^T|W^T}$ and attacker $p_{\hat{X}^T|Z^T}$ as RNNs, which are well-suited for the time-series structure of the SMs data and online processing. In the following, we describe in detail the loss functions, the network architectures and the learning algorithm used to train the releaser mechanism.

\subsection{Loss Functions}

Considering \eqref{eqtr2} and \eqref{eq:di_criterion}, the loss function for the releaser is defined as follows:
\begin{equation} \label{eq:releaser_loss} \mathcal{L}_{\mathcal{R}}(\theta, \phi,\lambda) \coloneqq D(Z^T,Y^T) - \frac{\lambda}{T} \sum_{t=1}^{T} H\big(\hat{X}_t|Z^t\big), \end{equation}
where $\lambda \ge 0$ controls the privacy-utility trade-off, $\theta$ are the parameters of the releaser and $\phi$ are the parameters of the adversary. It should be mentioned that for $\lambda = 0$, the loss function $\mathcal{L}_{\mathcal{R}}(\theta,\phi,\lambda)$ reduces to the expected distortion, being independent from the adversary. In such scenario, the releaser offers no privacy guarantees. Conversely, for very large values of $\lambda$, the loss function $\mathcal{L}_{\mathcal{R}}(\theta,\phi,\lambda)$ is dominated by the second term, so that privacy is the main goal of the releaser. In this regime, we expect the attacker to fail in inferring $X^T$, i.e., to approach to random guessing performance.\\

On the other hand, from \eqref{eq:attacker_optimization_problem}, the adversary loss function is defined as follows:% optimizes the following cross-entropy loss:
\begin{equation} \label{eq:attacker_loss} \mathcal{L}_{\mathcal{A}}(\phi) \coloneqq \frac{1}{T} \sum_{t=1}^T \E\left[- \log p_{\hat{X}_t|Z^t}(X_t|Z^t) \right], \end{equation}
where the expectation is with respect to $p_{X_t Z^t}$.

It should be mentioned that for training the previous loss functions are approximated by evaluating the expectations empirically as shown next. Let $\{(x^{(b)T},y^{(b)T})\}_{b=1}^{B}$ be a sample of $B$ examples and $\{z^{(b)T}\}_{b=1}^{B}$ the corresponding outputs of the releaser. Then, the loss functions are approximated as follows:
\begin{align} &\mathcal{L}_{\mathcal{R}}(\theta, \phi,\lambda)  \approx \frac{1}{BT} \sum_{b=1}^B d(z^{(b)T},y^{(b)T})  \nonumber \\ \label{eq:loss_R} & + \frac{\lambda}{BT}\sum_{b=1}^{B} \sum_{\hat{x}^{(b)}_t \in \mathcal{X}} p_{\hat{X}_t|Z^t}(\hat{x}^{(b)}_t|z^{(b)t}) \, \log p_{\hat{X}_t|Z^t}(\hat{x}^{(b)}_t|z^{(b)t}), \\ \mathcal{L}_{\mathcal{A}}(\phi) & \approx - \frac{1}{BT} \sum_{t=1}^T \sum_{b=1}^B \log p_{\hat{X}_t|Z^t}(x_t^{(b)}|z^{(b)t}).
\end{align}

\subsection{Recurrent Neural Networks and Long Short-Term Memory}

\begin{figure}[htbp]
	\centering
	\includegraphics[width=0.9\linewidth]{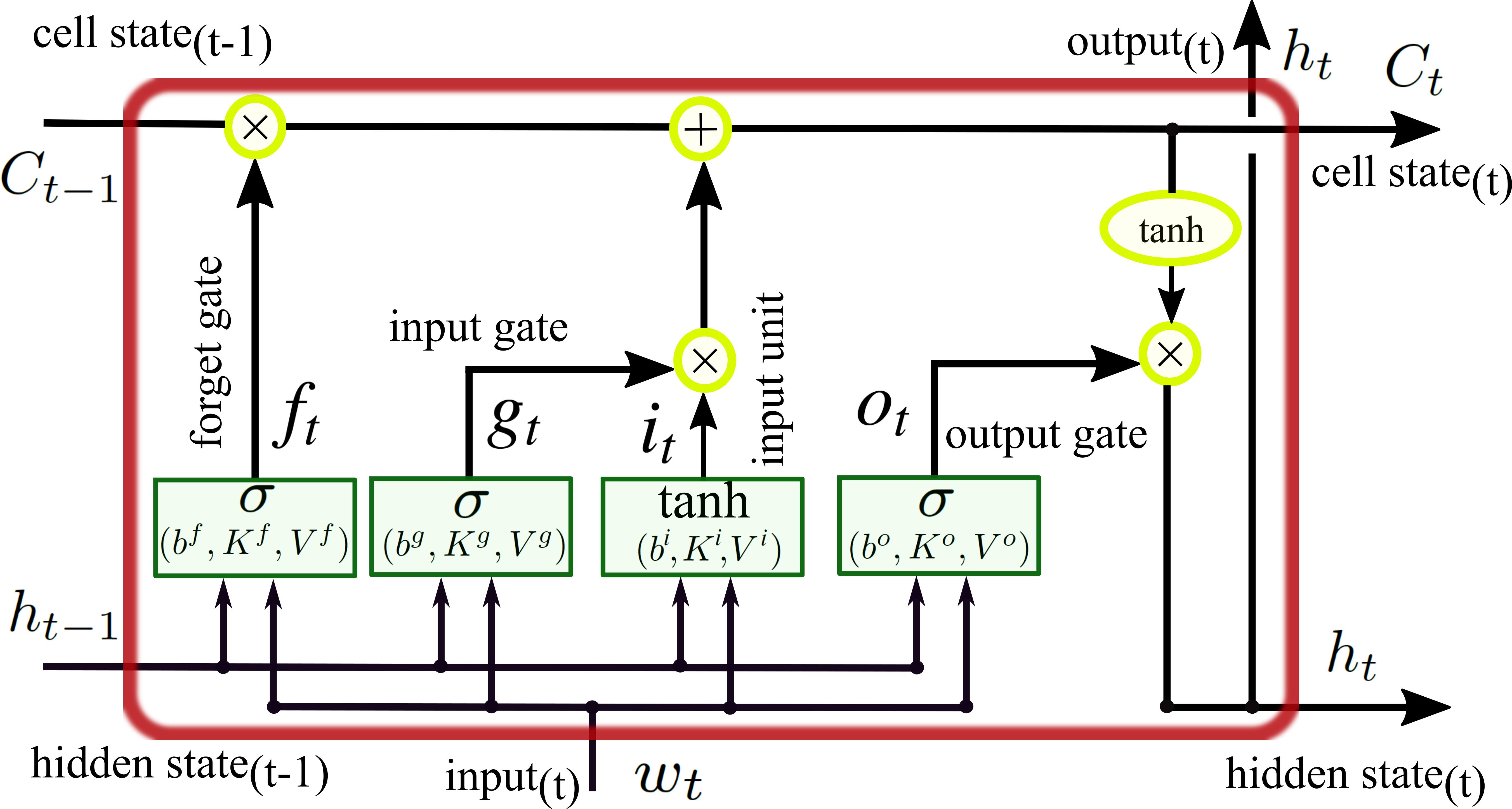}
	\caption{LSTM recurrent network cell diagram at time step $t$.}
	\label{had2}
\end{figure}

RNNs are a class of neural networks that are able to process sequential data by modeling the temporal correlation in data. Therefore, the output of an RNN network at time step $t-1$ generally affects the output at time $t$. Training of the RNNs is generally performed by gradient descent using the backpropagation through time algorithm \cite{werbos1990backpropagation}. However, learning long-term dependencies of time series data by RNNs may lead to the gradient vanishing or exploding problems, thus preventing successful training \cite{bengio1994learning}. To resolve this issue, the so-called LSTM cell was introduced in \cite{hochreiter1997long} and further improved in \cite{gers1999learning}. Fig. \ref{had2} represents the architecture of a LSTM cell in detail.

The LSTM cell includes four gating units to control the flow of information. All the gating units have a sigmoid activation function ($\sigma$) except for the input unit that uses an hyperbolic tangent activation function ($\tanh$) by default. Considering $w^T=(w_1,w_2,\dots,w_T)$ as the input, the formulations of the LSTM nodes at time step $t$ are represented in equation \eqref{LSTM_H} where the parameters $b,K,V$ are respectively biases, input weights, and recurrent weights:

\begin{equation}\label{LSTM_H}
\begin{aligned}
    f_t &= \sigma(b^f + K^fh_{t-1} + V^f w_t)\\
    g_t &= \sigma(b^g + K^g h_{t-1} + V^g w_t)\\
    i_t &= \sigma(b^i + K^i h_{t-1} + V^i w_t)\\
    o_t &= \sigma(b^o + K^oh_{t-1} + V^ow_t)\\
    C_t &= f_t\times C_{t-1} + g_t\times i_t\\
    h_t &= o_t\times \tanh\left(C_t\right)\\
\end{aligned}
\end{equation}

In the LSTM architecture, the forget gate $f_t $ uses the output of the previous cell (which is called hidden state $h_{t-1}$) to control the cell state $C_t$ to remove irrelevant information. On the other hand, the input gate $g_t$ and input unit $i_t$ adds new information to $C_t$ from the current input. Finally, the output gate $o_t$ generates the output of the cell from the current input and cell state. For more details on RNNs and LSTMs, the reader is referred to \cite{goodfellow2016} and references therein.

\subsection{Privacy-Preserving Adversarial Learning}

\begin{figure}[t!]
	\centering
	\includegraphics[width=1\linewidth]{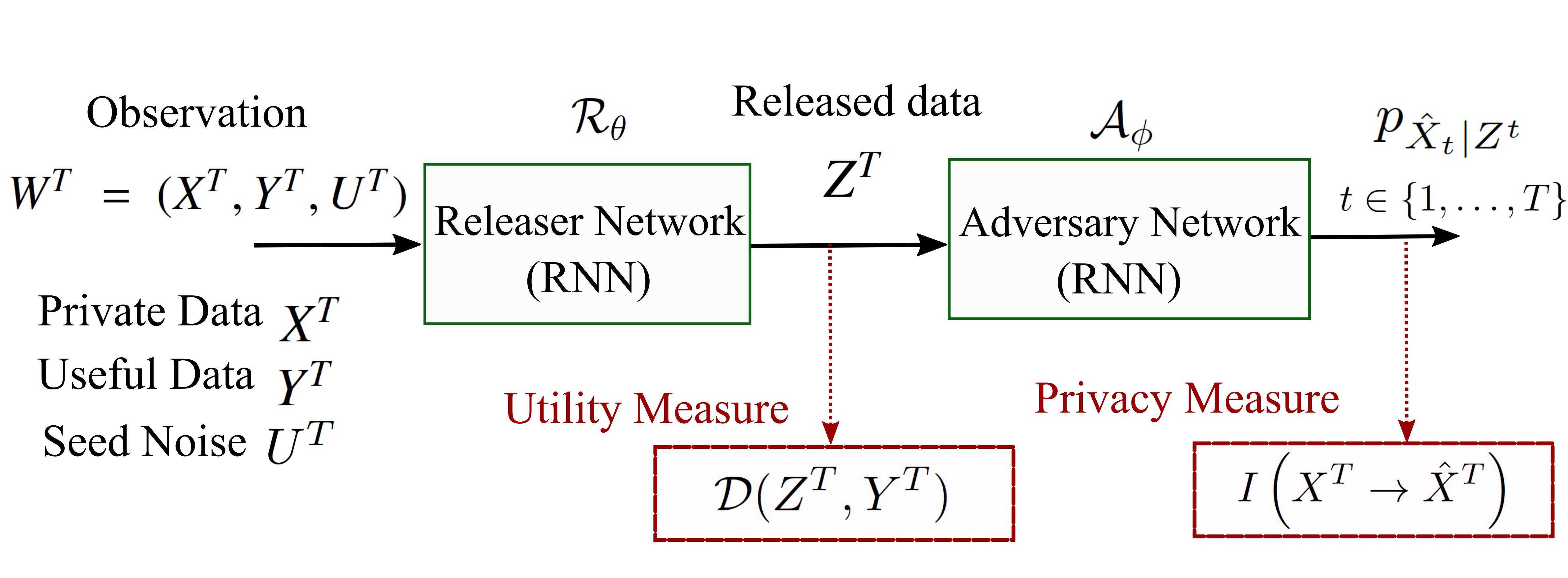}
	\caption{Privacy-Preserving framework. The seed noise $U^T$ is generated from i.i.d. samples according to a uniform distribution: $U_t \sim U[0,1]$.}
	\label{had1}
\end{figure}

Based on the previous formulation, an adversarial modeling framework consisting of two RNNs, a releaser $\mathcal{R}_{\theta}$  and an adversary $\mathcal{A}_{\phi}$, is considered (see Fig. \ref{had1}). Note that independent noise $U^T$ (with dimension $m$) is appended to $W^T$ in order to randomize the released variables $Z^T$, which is a popular approach in privacy-preserving methods. In addition, the available theoretical results show that, for Gaussian distributions, the optimal release contains such a noise component \cite{sankar2013smart,tripathy2019privacy}. For both networks, a LSTM architecture is selected. Training in the suggested framework is performed using the Algorithm \ref{Al1} which requires  $k$ gradient steps to train $\mathcal{A}_\phi$ followed by one gradient step to train $\mathcal{R}_\theta$. It is worth to emphasize that $k$ should be large enough in order to ensure that $\mathcal{A}_\phi$ represents a strong adversary. However, if $k$ is too large, this could lead to overfitting and thus a poor adversary. After the training of both networks is completed, a new network is trained from scratch in order to test the privacy achieved by the releaser network.

\begin{algorithm*}
	\caption{Algorithm for training privacy-preserving data releaser neural network.}
	\label{Al1}
	\textbf{Input:} Dataset (which includes samples of useful data $y^T$, sensitive data $x^t$); seed noise samples $u^T$; seed noise dimension $m$; batch size $B$; number of steps to apply to the Adversary $k$; gradient clipping value $C$; $\ell_2$ regularization parameter $\beta$. \\
	\textbf{Output:} Releaser network $\mathcal{R}_\theta$.
	\begin{algorithmic}[1]
		\FOR {number of training iterations}
	
		\FOR {$k$ steps}
		\STATE Sample minibatch of $B$ examples $\{ w^{(b)T} = (x^{(b)T},y^{(b)T},u^{(b)T})\}_{b=1}^B$ and generate releases $\{ z^{(b)T}\}_{b=1}^B$.% $\{ z^{(b)T}; \; b=1,2,\ldots,B \}$.
		\STATE Compute the gradient of $\mathcal{L}_{\mathcal{A}}(\phi)$, empirically approximated with the minibatch $\mathcal{B}$, with respect to $\phi$.
		\STATE Update $\phi$ by applying the RMSprop optimizer \cite{hinton2012neural} with clipping value $C$.
		\ENDFOR
		
		\STATE Sample minibatch of $B$ examples $\{ w^{(b)T} = (x^{(b)T},y^{(b)T},u^{(b)T})\}_{b=1}^B$ and generate releases $\{ z^{(b)T}\}_{b=1}^B$.
		
		\STATE Compute the gradient of $\mathcal{L}_{\mathcal{R}}(\theta,\phi,\lambda)$, approximated with the minibatch $\mathcal{B}$, with respect to $\theta$. 
		\STATE Use $\textrm{Ridge}(L_2)$ regularization \cite{hastie2005elements} with value $\beta$ and update $\theta$ by applying RMSprop optimizer with clipping value $C$.
		\ENDFOR
	\end{algorithmic}
\end{algorithm*}

\section{Results and Discussion} \label{sec:results}

We will validate our results on the Electricity Consumption \& Occupancy (ECO) dataset. ECO is collected and published by \cite{beckel2014eco}, which includes 1 Hz power consumption measurements and occupancy information of five houses in Switzerland over a period of $8$ months. Occupancy labels are determined as $1$ for the case that someone is at home and $0$ otherwise. Thus, for this application, the privacy attacker is a binary classifier that attempts to infer if a household is occupied or not at a given time. In this study, we re-sampled the data to have hourly samples. We model the time dependency over each day, so the dataset is reshaped to sample sequences of length $24$. A total number of $11225$ sample sequences were collected. The datasets are split into training and test sets with a ratio of roughly \mbox{85:15} while $10 \%$ of training data is dedicated to  validation which intended to set the hyperparameters. The network architectures and hyperparameters values are summarized in Table \ref{tab:hyperparameters}. A  stronger attacker composed of 3 LSTM layers is used for the test.

\begin{table*}[htbp]
	\centering
	\caption{Model architectures and hyperparameters values.}% used for inference of households occupancy application
	\begin{adjustbox}{width=0.7\textwidth}
		\begin{tabular}{c c c c c c c}
			\toprule
			%\midrule
			\textbf{\makecell{Releaser}}& \textbf{\makecell{Adversary}}& \textbf{\makecell{Attacker}}& \textbf{\makecell{$B$}}& \textbf{\makecell{$k$}}& \textbf{\makecell{$m$}}\\ 
			\midrule[0.5pt]
			\makecell{$4$ LSTM layers each\\with 64 cells and $\beta=1.5$} &\makecell{$2$ LSTM layers each\\ with 32 cells}& \makecell{$3$ LSTM layers each\\ with 32 cells}&128 &4&8 \\
		
			\midrule
			\bottomrule
			
		\end{tabular}
		%\centering
	\end{adjustbox}
	\label{tab:hyperparameters}
\end{table*}

To clearly assess the distortion with respect to the actual power consumption measurements, we define the Normalized Error (NE) for the different $\ell_p$ distortion functions as follows:

\begin{equation} \text{NE}_p \coloneqq \frac{\E\left[ \| Y^T - Z^T \|_p \right]}{\E\left[ \| Y^T \|_p \right]}. \end{equation}
In addition, performance of the attacker on inferring the private attributes is quantified based on the balanced accuracy. This is common in classification problems to deal with the data imbalance problem, which occurs when the number of samples for each class is quite different. Balanced accuracy is defined as the average recall calculated for each class \cite{mosley2013balanced}. Concretely, let $c_{ij}$ represent the fraction of examples of class $i$ classified as class $j$. Then, the balanced accuracy can be defined as
\begin{equation} \text{Balanced Accuracy} \coloneqq \frac{1}{2} \left( \frac{c_{11}}{c_{11}+c_{12}} + \frac{c_{22}}{c_{22}+c_{21}} \right). \end{equation}
This metric provides a fair assessment of the quality of the attacker independently from the degree of data unbalance. Thus, simplifying the analysis of the results. In the following we use the term accuracy to refer to the balanced accuracy.

\subsection{$\ell_2$ Distortion} \label{subsec:L2norm}

\begin{figure}[htbp]
	\centering
	\includegraphics[width=.6\linewidth]{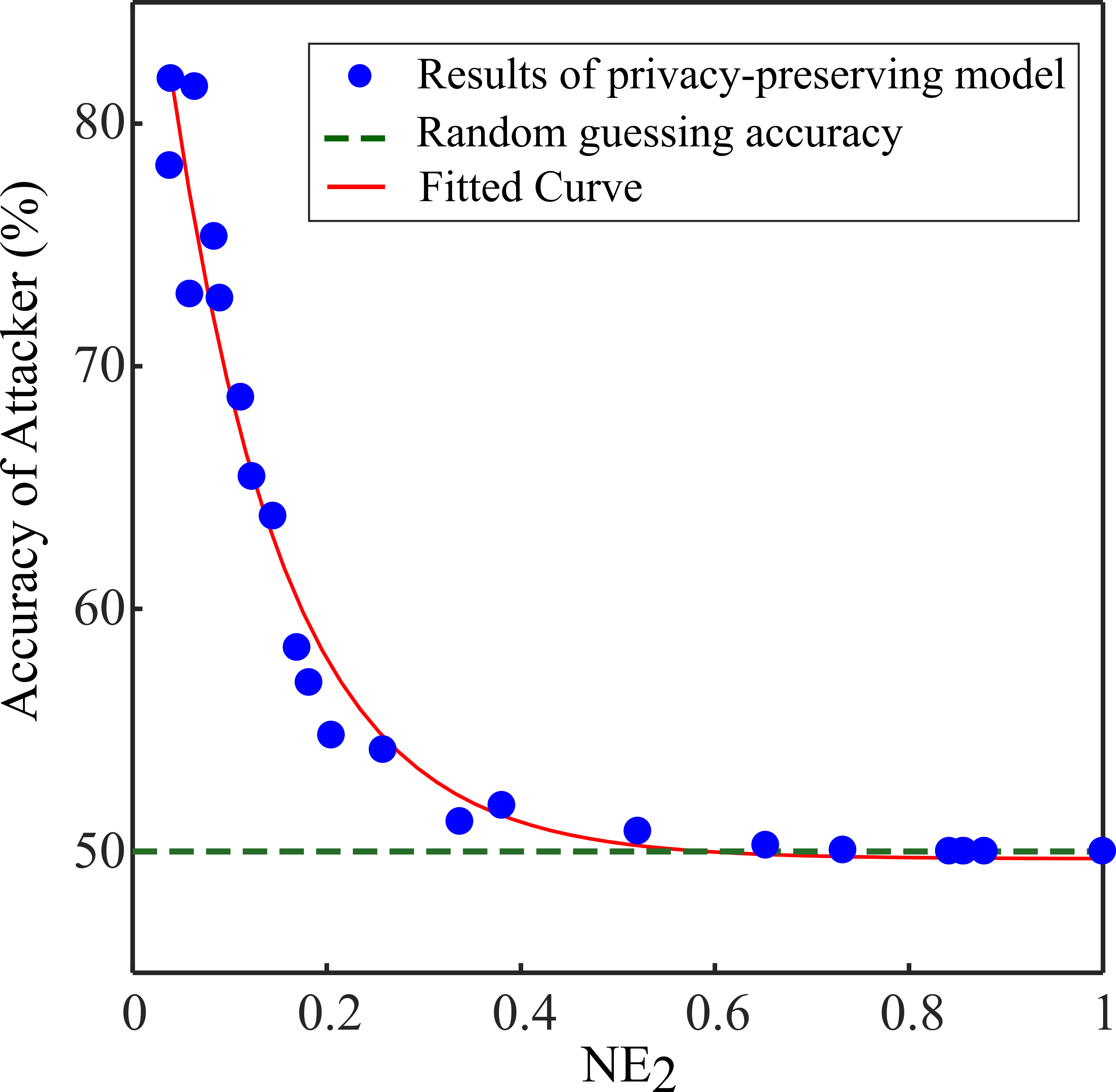}
	\caption{Privacy-utility trade-off for house occupancy inference using $\ell_2$ distortion function. The fitted curve is based on an exponential function and is included only for illustration purposes.}
	\label{ECOTradeoff}
\end{figure}

In this section, we consider the $\ell_2$ distortion function (i.e., $p = 2$ in \eqref{eq:p_distortion}). Fig. \ref{ECOTradeoff} shows the empirically found privacy-utility trade-off for this scenario. Note that by increasing the distortion of the release, the accuracy of the attacker changes from more than $80 \%$ (almost no privacy) to $50 \%$ (full privacy).

To assess the quality of the release signal, utility providers may be interested in several different indicators. These include, for instance, the mean, skewness, kurtosis, standard deviation to mean ratio, and maximum to mean ratio \cite{shen2019}. Thus, for completeness, we present these indicators in Table \ref{tab:tab1} for three different cases along the privacy-utility trade-off curve. We observe that in general the errors in these indicators are small when the privacy constraints are lax and increase as they become stricter. Nevertheless, no simple relation can be expected between NE$_2$ and the values of those indicators.\\

\begin{table*}
	\centering
	\caption{Errors in power quality indicators along the privacy-utility trade-off.}
	\begin{adjustbox}{width=0.7\textwidth}
		\renewcommand{\arraystretch}{0.5}
		\begin{tabular}{c c c c c c c}
			\toprule
			\midrule
			\textbf{NE$_2$} & \textbf{Accuracy(\%)} & \multicolumn{5}{c}{\textbf{Absolute relative error of quality indicators(\%) } }\\ 
			\cmidrule(rl){3-7}
			& & \textit{Mean}  & \textit{Skewness}& \textit{Kurtosis}& \textit{Std. Dev./Mean}& \textit{Max./Mean}  \\ 
			\midrule
			0.04& 78 &1.42&1.06&0.70&0.67&0.46\\
			0.12&65&9.69&4.32&5.81&4.58&4.92\\
			0.18&57& 13.26&12.83&2.57&16.44&13.89\\
			\midrule
			\bottomrule
			
		\end{tabular}
		%\centering
	\end{adjustbox}
	\label{tab:tab1}
\end{table*}
%\normalsize

\subsubsection{Comparison with regular random noise addition approach}

\quad\\

\begin{figure}[htbp]
	\centering
	\includegraphics[width=0.98\linewidth]{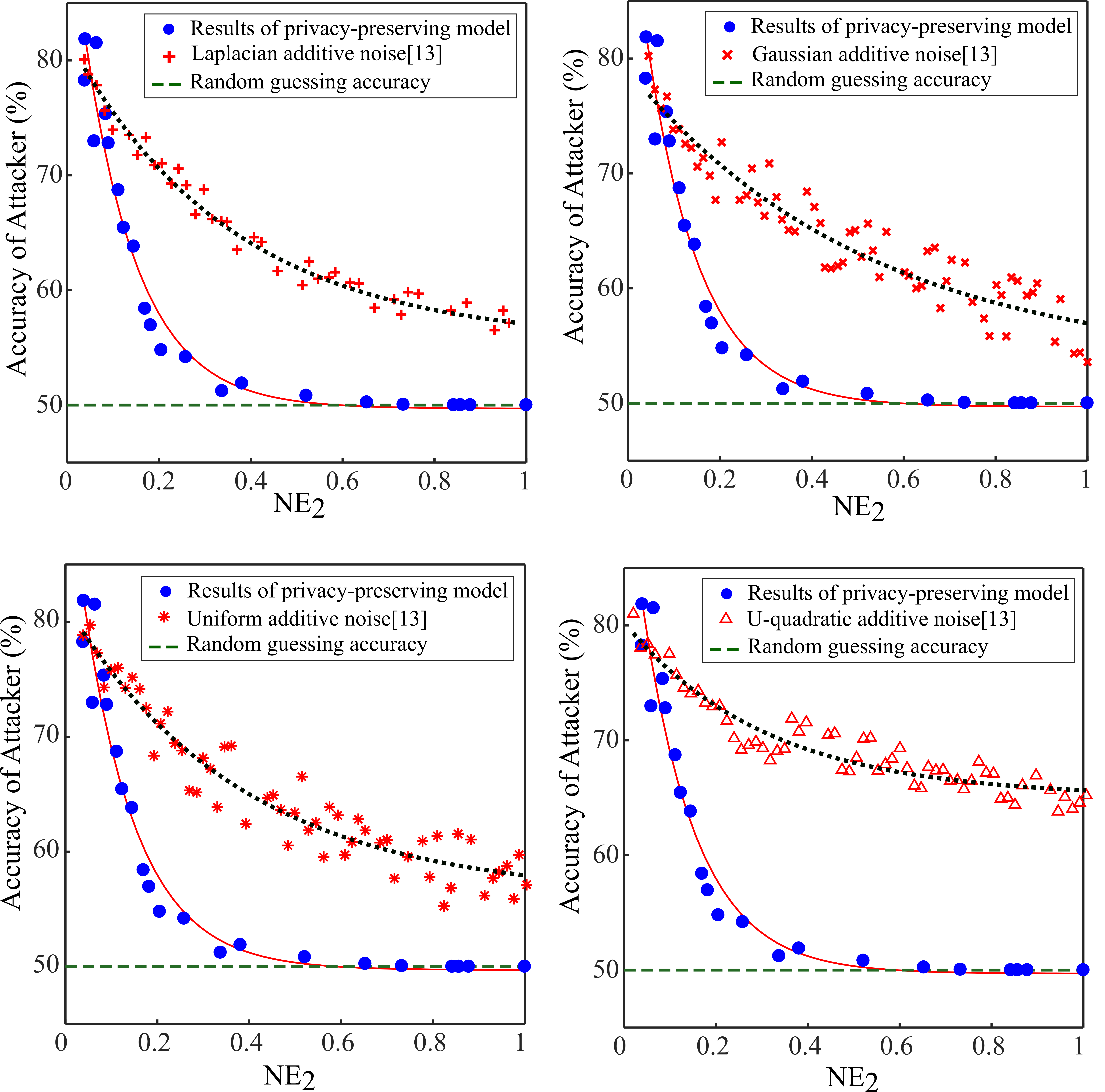}
	\caption{Privacy-utility trade-off for house occupancy inference using the random noise addition approach.}
	\label{random_noise_tradeoff}
\end{figure}

As it was discussed in Section III, the proposed model in this study provides privacy through distorting the SMs data. However, in contrast with the regular random noise addition approaches~\cite{barbosa2016} where a random noise $E_t$ is added to the SMs data (i.e. $Z_t = Y_t + E_t,$ for $t\in\{1,2,\dots,T\}$), our model distorts the SMs data by performing a noisy recurrent transformation on $Z^T$. To compare our method with the regular random noise addition method, four different cases for random noise $E_t$ were considered: Laplacian, Gaussian, Uniform, and U-quadratic. In all cases, the amount of distortion is controlled by the variance of noise. The same type of attacker as the one presented in Table \ref{tab:hyperparameters} is used to infer the private data out of the distorted one. Fig.~\ref{random_noise_tradeoff} shows the privacy-utility trade-off for these random noise addition approaches as compared with our model. It can be seen clearly that, for the same amount of distortion, our method is more successful in preventing the attacker from inferring the private information. This is expected, as our method is able to learn the noise distribution to fit the actual demand load and the sensitive information that is being hidden from the attacker, which is a much more powerful approach than just using a fixed noise distribution. Notice also that, as expected, all results are similar as the distortion approaches to zero. For some of the random noise addition cases, the fitted curves cross over the one of our model, but in the low distortion low privacy area of the graph, which is not of interest.

\subsubsection{Comparison with PPAN model}

\quad\\

\begin{figure}[htbp]
	\centering
	\includegraphics[width=0.6\linewidth]{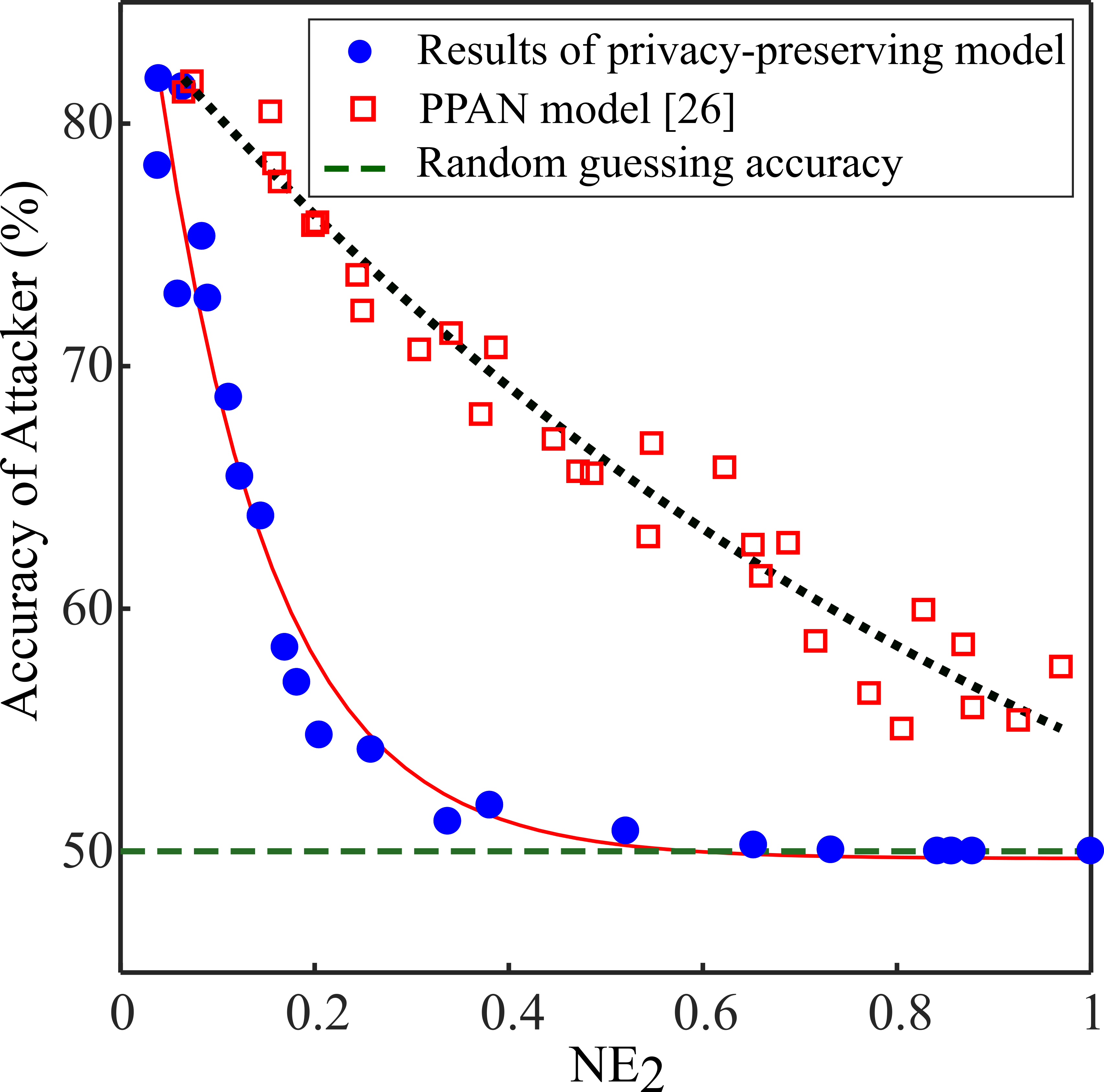}
	\caption{Privacy-utility trade-off for house occupancy inference using the PPAN approach.}
	\label{ppan_tradeoff}
\end{figure}

The main limitation of the random noise addition approach is that the noise mechanism is independent of the SMs data. A more sophisticated strategy that address this issue would be the PPAN model~\cite{tripathy2019privacy}. This method uses a variational lower bound on $I(X_t;Z_t)$ to train a releaser using an adversarial learning approach. For more details of this approach the readers are referred to~\cite{tripathy2019privacy}. Fig.~\ref{ppan_tradeoff} shows the privacy-utility trade-off of the PPAN model as compared with our method using the same type of attacker as presented in Table \ref{tab:hyperparameters}. For the PPAN, both the mechanism and adversary networks are deep neural networks including three hidden layers with 64 nodes and a rectified linear unit (ReLU) activation function, while the RMSprop optimizer~\cite{tieleman2012lecture} with learning rate 0.01 is used.

From Fig.~\ref{ppan_tradeoff} it can be seen that our method can clearly outperform the PPAN. The reason for this gap can be attributed to both the different cost function formulation (i.e., the fact that we are using an upper bound on $I(X^T \to \hat{X}^T)$ instead of a lower bound on $I(X_t;Z_t)$) and the recurrent structure of the proposed releaser mechanism.

\subsection{$\ell_p$ Distortion}% with $p>2$}

As already discussed in Section \ref{sec:formulation}, the distortion function should be properly matched to the intended application of the released variables $Z^T$ in order to preserve the characteristics of the target variables $Y^T$ that are considered essential. In this section, we consider the $\ell_p$ distortion \eqref{eq:p_distortion} with $p=4,5$ as an alternative to the $\ell_2$ distortion function used in the previous section and study their potential benefits.

\begin{figure}[htbp]
	\centering
	\includegraphics[width=1\linewidth]{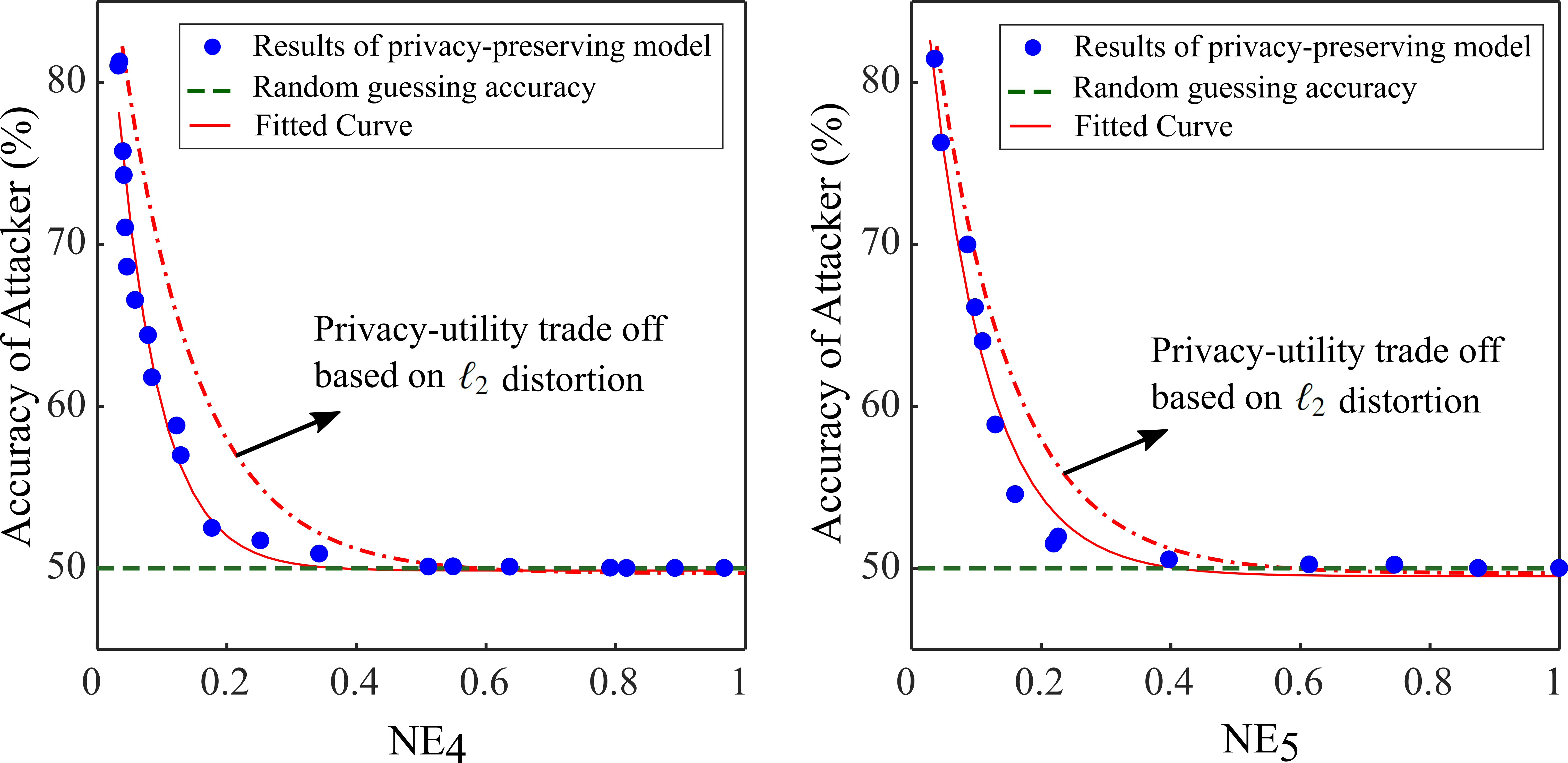}
	\caption{Privacy-utility trade-off for house occupancy inference based on the different $\ell_p$ distortion functions. For each figure, the dashed line, shown for comparison purposes, is the fitted curve found in Fig. \ref{ECOTradeoff} for the $\ell_2$ distortion function.}
	\label{LossesEco}
\end{figure}

The privacy-utility trade-off curve for these distortion functions is shown in Fig. \ref{LossesEco}. As a first observation, it is clear that the choice of the distortion measure has a non-negligible impact on the privacy-utility trade-off curve. In fact, it can be seen that for a given amount of normalized distortion, the releaser trained with the $\ell_4$ and $\ell_5$ distortion measures achieve a higher level of privacy than the one trained with the $\ell_2$ distortion function. It should also be mentioned that we also considered other norms, such as the $\ell_{10}$, and the privacy-utility trade-off was observed to be similar, but slightly different, than the one corresponding to the $\ell_4$ norm.

\begin{figure}[htbp]
	\centering
	\includegraphics[width=1\linewidth]{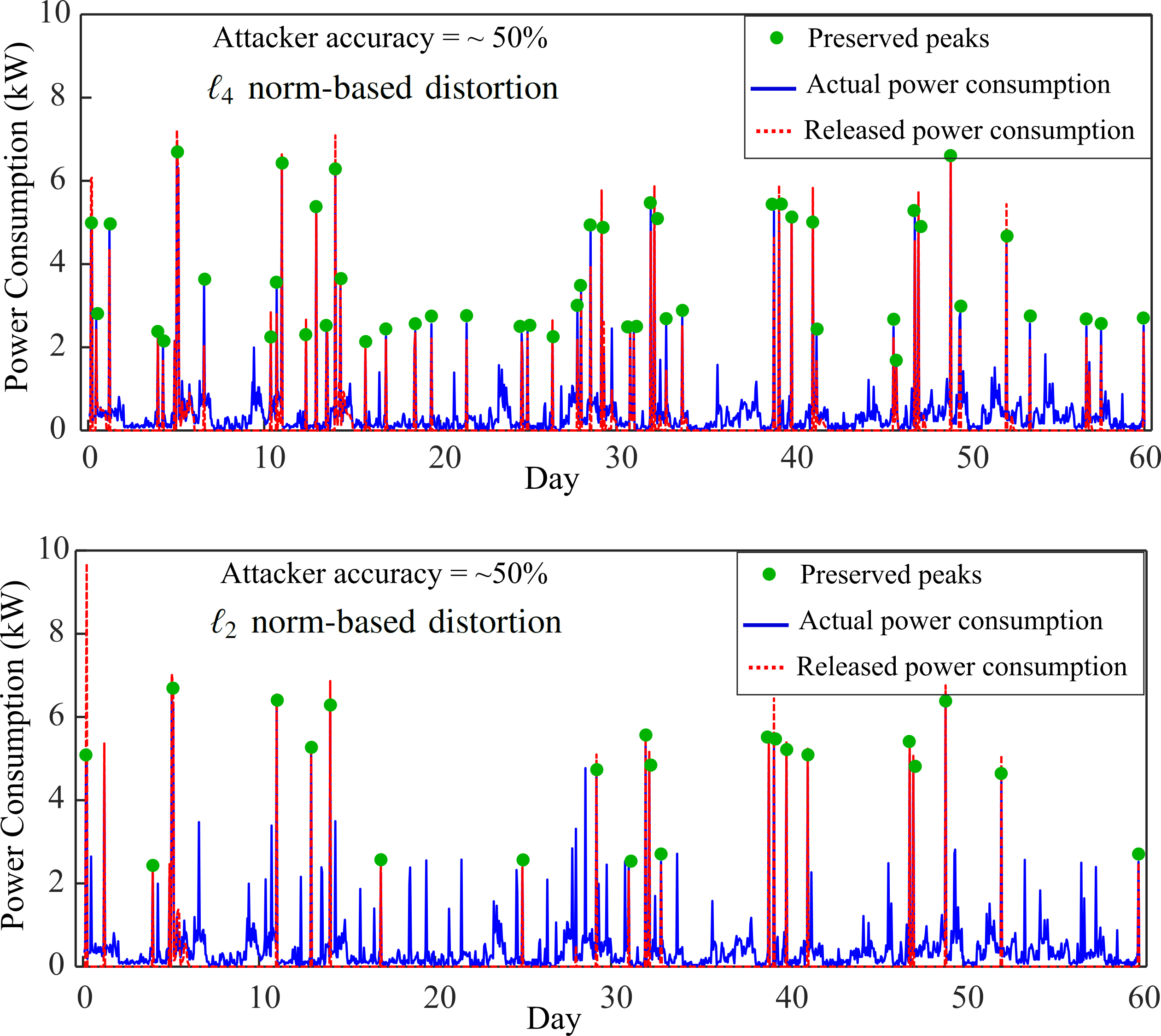}
	\caption{Example of the release power consumption in the time domain compared with the actual power consumption over 60 random days with almost full privacy for the $\ell_4$ and $\ell_2$ distortion functions.}
	\label{TimeL4L2ECO}
\end{figure}

As we discussed in Section \ref{sec:formulation}, in some applications, such as demand response programs, the utilities are mostly interested in the peak power consumption of the customers. It is also expected that higher-order $\ell_p$ norms are better at preserving these signal characteristics than the $\ell_2$ norm. To verify this notion, we considered 60 random days of the ECO dataset in a full privacy scenario (i.e., with an attacker accuracy very close to $50 \%$) and plotted the actual power consumption along with the corresponding release signals for both the $\ell_4$ and $\ell_2$ distortion functions. Results shown in Fig. \ref{TimeL4L2ECO} clearly indicate that the number of peaks preserved by the releaser trained with the $\ell_4$ distortion function is much higher than the ones kept by the releaser trained with the $\ell_2$ distortion function. This suggests that for these applications, higher order $\ell_p$ distortion functions should be considered. %In general, one can think of different release mechanisms tailored 

\subsection{Attacker with Data Mismatch Problem}

\begin{figure}[htbp]
	\centering
	\includegraphics[width=.7\linewidth]{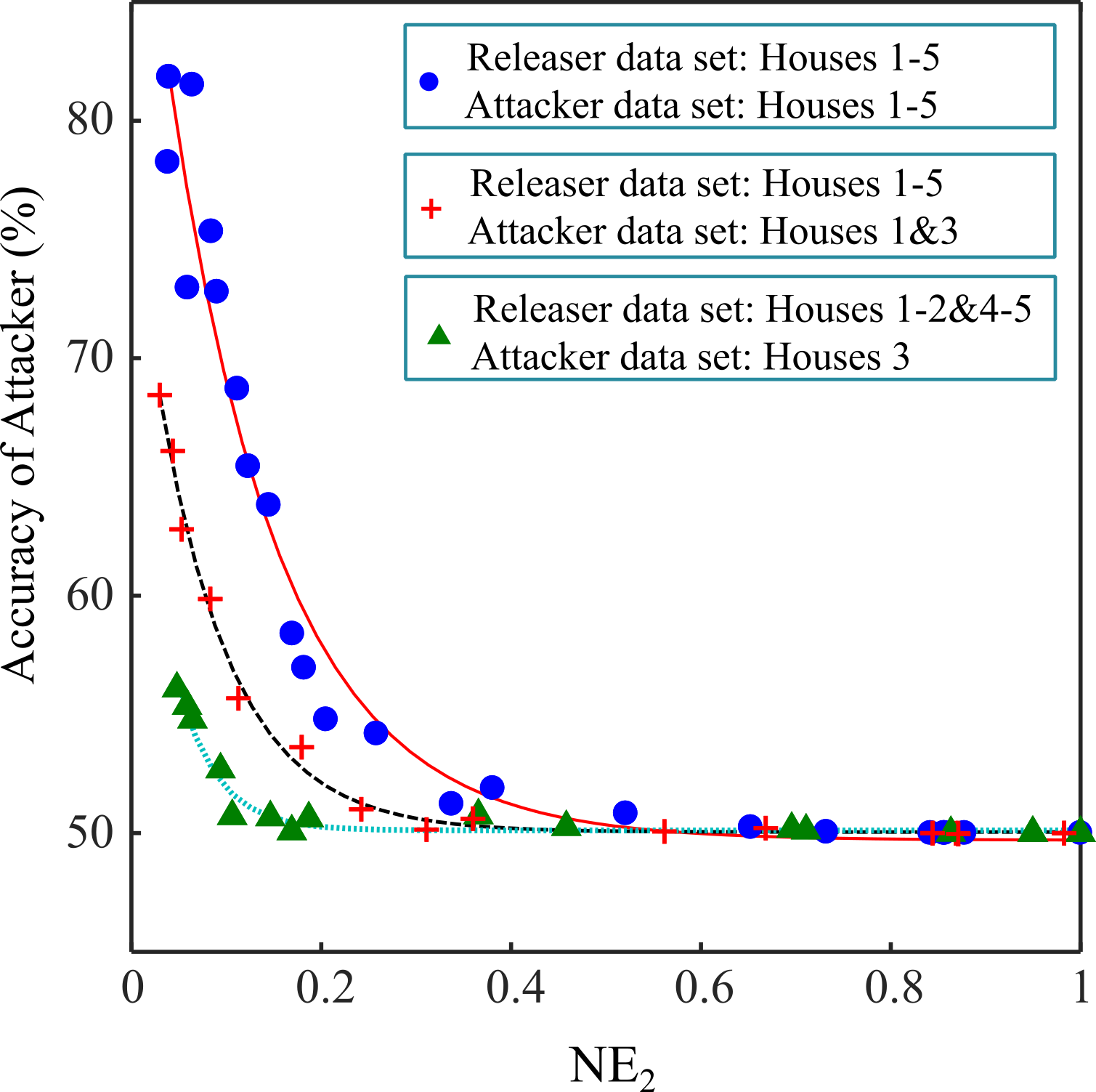}
	\caption{Effect of data mismatch between the releaser and the attacker on the privacy-utility trade-off for house occupancy inference.}
	\label{DiffDataset}
\end{figure}

All the previous results are based on the assumption that the attacker has access to exactly the same training dataset used by the releaser-adversary system. This case should be considered as a worst-case analysis of the performance of the releaser. However, this assumption may not be true in practice. To examine the impact of this hypothesis, we consider two different cases. It should be noted that the total number of samples used for training and testing was kept fixed in all the different scenarios. In the first case, we assume that, out of the dataset of the five houses in the ECO dataset, the releaser uses the data of all the houses for training while the attacker has only access to the data of houses $1$ and $3$. In the second case, we assume that releaser is trained with the data of houses $\{1,2,4,5\}$ but the attacker has only access to data from house $3$. These scenarios try to capture different degrees of the data mismatch problem, which could have an impact on the privacy-utility trade-off due to the different generalization errors. The results are presented in Fig. \ref{DiffDataset} along with the worst-case scenario. We conclude that the overlapping of the training datasets of the releaser and the attacker can strongly affect the performance of the model. In fact, in the case where the attacker does not have access to the same dataset as the releaser, its performance largely degrades, which means that a target level of privacy requires much less distortion. In the extreme case where the attacker has no access to the releaser training dataset, a very high level of privacy can be achieved with negligible distortion. It should be mentioned that we repeated this experiment with different choices of these 5 houses and similar results were obtained.

\section{Discussion and Concluding Remarks} \label{sec:conclusion}

Privacy concerns associated with SMs data-sharing are an important problem  since these can have an impact on their deployment pace and the advancement of smart grid technologies. As a consequence, it is essential to understand and to palliate real privacy risks in order to provide an adequate solution to conveniently share SMs data. A summary of the privacy-aware SMs methodology proposed in this study and the key findings are provided below:

\begin{itemize}
	
	\item We proposed to measure the privacy based on the DI between the sensitive time series and its inference by a potential attacker optimized for the dedicated task. This captures the causal time dependencies present in the time series data and its sequential processing. For the sake of computational tractability, we propose an upper bound to the DI which leads to our training objective. Then, using Fano's inequality, it was shown that this bound can constrain the performance of the optimal (Bayesian) attacker. 
	
	\item We explored a data release framework that balances the trade-off between privacy of the sensitive information and distortion of the useful data. The desired releaser was trained using RNNs in an adversarial framework to optimize such objective, while an adversary mimics a real attacker. After convergence, an actual attacker was trained to test the level of privacy achieved by the releaser.
	
	\item A detailed study of the inference of households occupancy using actual SMs dataset was performed. The performance of the proposed model was compared with two methodologies: random noise addition and PPAN-based. In both cases, the results show that our method clearly outperforms the other algorithms in limiting the attacker inference ability. This is not surprising, as our method is able to fit the noise distribution to the actual consumer load and the sensitive feature that is trying to be hidden by the releaser in a more effective manner.
	
	\item We showed that the choice of the distortion measure can have a significant impact on the privacy-utility trade-off curve. Indeed, it is shown that the $\ell_4$ distortion measure generates a release that preserves most of the power consumption peaks even under a full privacy regime, which is not the case for the $\ell_2$ distortion function. This result may be of considerable importance for some applications such as demand response. More generally, our formulation is amenable to train different release systems tailored for several potential applications based on SMs data.

	\item We studied the impact of the data mismatch problem in this application, which occurs when the training dataset of the attacker is not exactly the same as the one used by the releaser. These results have shown that this effect can greatly affect the privacy-utility trade-off. Since this phenomenon is expected in practice, at least to some degree, these findings suggest that the level of required distortion to achieve desired privacy targets may be negligible in several cases of interest.

\end{itemize}{}

To wrap up the paper, two possible extensions for future work are briefly discussed. The first one is modeling the availability of side information at the attacker side to model prior knowledge of sensitive information as in \cite{salehkalaibar2019}, which cannot be distorted by the releaser, and study its impact on the privacy-utility trade-offs. The second is combining this SM data distortion approach with the ones which rely on physical resources for demand shaping. This would require us to incorporate the electricity cost consideration, leading to the study of the privacy-utility-cost trade-offs.

\section*{Acknowledgment}
This work was supported by Hydro-Quebec, the Natural Sciences and Engineering Research Council of Canada, and McGill University in the framework of the NSERC/Hydro-Quebec Industrial Research Chair in Interactive Information Infrastructure for the Power Grid (IRCPJ406021-14). This project has received funding from the European Union’s Horizon 2020 research and innovation programme under the Marie Skłodowska-Curie grant agreement No 792464.%The work of Prof. Pablo Piantanida was supported by the European Commission’s Marie Sklodowska-Curie Actions (MSCA), through the Marie Sklodowska-Curie IF (H2020-MSCAIF-2017-EF-797805-STRUDEL).

\bibliographystyle{ieeetr}
\bibliography{main}

\begin{thebibliography}{10}

\bibitem{alahakoon2016}
D.~{Alahakoon} and X.~{Yu}, ``Smart electricity meter data intelligence for
  future energy systems: A survey,'' {\em IEEE Transactions on Industrial
  Informatics}, vol.~12, pp.~425--436, Feb 2016.

\bibitem{wang2019}
Y.~{Wang}, Q.~{Chen}, T.~{Hong}, and C.~{Kang}, ``Review of smart meter data
  analytics: Applications, methodologies, and challenges,'' {\em IEEE
  Transactions on Smart Grid}, vol.~10, pp.~3125--3148, May 2019.

\bibitem{molina2010}
A.~Molina-Markham, P.~Shenoy, K.~Fu, E.~Cecchet, and D.~Irwin, ``Private
  memoirs of a smart meter,'' in {\em Proceedings of the 2Nd ACM Workshop on
  Embedded Sensing Systems for Energy-Efficiency in Building}, BuildSys '10,
  (New York, NY, USA), pp.~61--66, ACM, 2010.

\bibitem{mckenna2012}
E.~McKenna, I.~Richardson, and M.~Thomson, ``Smart meter data: Balancing
  consumer privacy concerns with legitimate applications,'' {\em Energy
  Policy}, vol.~41, pp.~807 -- 814, 2012.
\newblock Modeling Transport (Energy) Demand and Policies.

\bibitem{jain2016}
P.~Jain, M.~Gyanchandani, and N.~Khare, ``Big data privacy: a technological
  perspective and review,'' {\em Journal of Big Data}, vol.~3, p.~25, Nov 2016.

\bibitem{asghar2017}
M.~R. {Asghar}, G.~{Dán}, D.~{Miorandi}, and I.~{Chlamtac}, ``Smart meter data
  privacy: A survey,'' {\em IEEE Communications Surveys Tutorials}, vol.~19,
  pp.~2820--2835, Fourthquarter 2017.

\bibitem{giaconi2018privacy}
G.~Giaconi, D.~Gunduz, and H.~V. Poor, ``Privacy-aware smart metering: Progress
  and challenges,'' {\em IEEE Signal Processing Magazine}, vol.~35, no.~6,
  pp.~59--78, 2018.

\bibitem{li2010}
F.~{Li}, B.~{Luo}, and P.~{Liu}, ``Secure information aggregation for smart
  grids using homomorphic encryption,'' in {\em 2010 First IEEE International
  Conference on Smart Grid Communications}, pp.~327--332, Oct 2010.

\bibitem{rottondi2013}
C.~{Rottondi}, G.~{Verticale}, and C.~{Krauss}, ``Distributed
  privacy-preserving aggregation of metering data in smart grids,'' {\em IEEE
  Journal on Selected Areas in Communications}, vol.~31, pp.~1342--1354, July
  2013.

\bibitem{efthymiou2010smart}
C.~Efthymiou and G.~Kalogridis, ``Smart grid privacy via anonymization of smart
  metering data,'' in {\em 2010 First IEEE International Conference on Smart
  Grid Communications}, pp.~238--243, IEEE, 2010.

\bibitem{mashima2015authenticated}
D.~Mashima, ``Authenticated down-sampling for privacy-preserving energy usage
  data sharing,'' in {\em 2015 IEEE International Conference on Smart Grid
  Communications (SmartGridComm)}, pp.~605--610, IEEE, 2015.

\bibitem{eibl2015}
G.~{Eibl} and D.~{Engel}, ``Influence of data granularity on smart meter
  privacy,'' {\em IEEE Transactions on Smart Grid}, vol.~6, pp.~930--939, March
  2015.

\bibitem{barbosa2016}
P.~Barbosa, A.~Brito, and H.~Almeida, ``A technique to provide differential
  privacy for appliance usage in smart metering,'' {\em Information Sciences},
  vol.~370-371, pp.~355 -- 367, 2016.

\bibitem{sankar2013smart}
L.~{Sankar}, S.~R. {Rajagopalan}, S.~{Mohajer}, and H.~V. {Poor}, ``Smart meter
  privacy: A theoretical framework,'' {\em IEEE Transactions on Smart Grid},
  vol.~4, pp.~837--846, June 2013.

\bibitem{cover2006elements}
T.~M. Cover and J.~A. Thomas, ``Elements of information theory, 2nd edition,''
  {\em Willey-Interscience: NJ}, 2006.

\bibitem{zheng2017privacy}
P.~Zheng, B.~Chen, X.~Lu, and X.~Zhou, ``Privacy-utility trade-off for smart
  meter data considering tracing household power usage,'' in {\em 2017 IEEE 2nd
  Information Technology, Networking, Electronic and Automation Control
  Conference (ITNEC)}, pp.~939--943, IEEE, 2017.

\bibitem{kalogridis2010}
G.~{Kalogridis}, C.~{Efthymiou}, S.~Z. {Denic}, T.~A. {Lewis}, and R.~{Cepeda},
  ``Privacy for smart meters: Towards undetectable appliance load signatures,''
  in {\em 2010 First IEEE International Conference on Smart Grid
  Communications}, pp.~232--237, Oct 2010.

\bibitem{tan2013}
O.~{Tan}, D.~{Gunduz}, and H.~V. {Poor}, ``Increasing smart meter privacy
  through energy harvesting and storage devices,'' {\em IEEE Journal on
  Selected Areas in Communications}, vol.~31, pp.~1331--1341, July 2013.

\bibitem{zhang2016cost}
Z.~Zhang, Z.~Qin, L.~Zhu, J.~Weng, and K.~Ren, ``Cost-friendly differential
  privacy for smart meters: Exploiting the dual roles of the noise,'' {\em IEEE
  Transactions on Smart Grid}, vol.~8, no.~2, pp.~619--626, 2016.

\bibitem{sun2017smart}
Y.~Sun, L.~Lampe, and V.~W. Wong, ``Smart meter privacy: Exploiting the
  potential of household energy storage units,'' {\em IEEE Internet of Things
  Journal}, vol.~5, no.~1, pp.~69--78, 2017.

\bibitem{li2018information}
S.~Li, A.~Khisti, and A.~Mahajan, ``Information-theoretic privacy for smart
  metering systems with a rechargeable battery,'' {\em IEEE Transactions on
  Information Theory}, vol.~64, no.~5, pp.~3679--3695, 2018.

\bibitem{giaconi2018}
G.~{Giaconi}, D.~{Gündüz}, and H.~V. {Poor}, ``Smart meter privacy with
  renewable energy and an energy storage device,'' {\em IEEE Transactions on
  Information Forensics and Security}, vol.~13, pp.~129--142, Jan 2018.

\bibitem{wang2017}
Y.~{Wang}, N.~{Raval}, P.~{Ishwar}, M.~{Hattori}, T.~{Hirano}, N.~{Matsuda},
  and R.~{Shimizu}, ``On methods for privacy-preserving energy
  disaggregation,'' in {\em 2017 IEEE International Conference on Acoustics,
  Speech and Signal Processing (ICASSP)}, pp.~6404--6408, March 2017.

\bibitem{goodfellow2014}
I.~Goodfellow, J.~Pouget-Abadie, M.~Mirza, B.~Xu, D.~Warde-Farley, S.~Ozair,
  A.~Courville, and Y.~Bengio, ``Generative adversarial nets,'' in {\em
  Advances in Neural Information Processing Systems 27} (Z.~Ghahramani,
  M.~Welling, C.~Cortes, N.~D. Lawrence, and K.~Q. Weinberger, eds.),
  pp.~2672--2680, Curran Associates, Inc., 2014.

\bibitem{huang2018}
C.~Huang, P.~Kairouz, X.~Chen, L.~Sankar, and R.~Rajagopal, ``Generative
  adversarial privacy,'' {\em CoRR}, vol.~abs/1807.05306, 2018.

\bibitem{tripathy2019privacy}
A.~Tripathy, Y.~Wang, and P.~Ishwar, ``Privacy-preserving adversarial
  networks,'' in {\em 2019 57th Annual Allerton Conference on Communication,
  Control, and Computing (Allerton)}, pp.~495--505, IEEE, 2019.

\bibitem{feutry2018}
C.~{Feutry}, P.~{Piantanida}, Y.~{Bengio}, and P.~{Duhamel}, ``{Learning
  Anonymized Representations with Adversarial Neural Networks},'' {\em arXiv
  e-prints}, p.~arXiv:1802.09386, Feb 2018.

\bibitem{esteban2018}
C.~Esteban, S.~L. Hyland, and G.~R{\"a}tsch, ``Real-valued (medical) time
  series generation with recurrent conditional gans,'' {\em CoRR},
  vol.~abs/1706.02633, 2018.

\bibitem{shateri2019deep}
M.~Shateri, F.~Messina, P.~Piantanida, and F.~Labeau, ``Deep directed
  information-based learning for privacy-preserving smart meter data release,''
  in {\em 2019 IEEE International Conference on Communications, Control, and
  Computing Technologies for Smart Grids (SmartGridComm)}, pp.~1--7, IEEE,
  2019.

\bibitem{erdogdu2015}
M.~A. {Erdogdu} and N.~{Fawaz}, ``Privacy-utility trade-off under continual
  observation,'' in {\em 2015 IEEE International Symposium on Information
  Theory (ISIT)}, pp.~1801--1805, June 2015.

\bibitem{li2019}
Z.~{Li}, T.~J. {Oechtering}, and D.~{Gündüz}, ``Privacy against a hypothesis
  testing adversary,'' {\em IEEE Transactions on Information Forensics and
  Security}, vol.~14, pp.~1567--1581, June 2019.

\bibitem{massey1990causality}
J.~Massey, ``Causality, feedback and directed information,'' in {\em Proc. Int.
  Symp. Inf. Theory Applic.(ISITA-90)}, pp.~303--305, Citeseer, 1990.

\bibitem{ho2010}
S.~{Ho} and S.~{Verdu}, ``On the interplay between conditional entropy and
  error probability,'' {\em IEEE Transactions on Information Theory}, vol.~56,
  pp.~5930--5942, Dec 2010.

\bibitem{werbos1990backpropagation}
P.~J. Werbos {\em et~al.}, ``Backpropagation through time: what it does and how
  to do it,'' {\em Proceedings of the IEEE}, vol.~78, no.~10, pp.~1550--1560,
  1990.

\bibitem{bengio1994learning}
Y.~Bengio, P.~Simard, P.~Frasconi, {\em et~al.}, ``Learning long-term
  dependencies with gradient descent is difficult,'' {\em IEEE transactions on
  neural networks}, vol.~5, no.~2, pp.~157--166, 1994.

\bibitem{hochreiter1997long}
S.~Hochreiter and J.~Schmidhuber, ``Long short-term memory,'' {\em Neural
  computation}, vol.~9, no.~8, pp.~1735--1780, 1997.

\bibitem{gers1999learning}
F.~A. {Gers}, J.~{Schmidhuber}, and F.~{Cummins}, ``Learning to forget:
  continual prediction with lstm,'' in {\em 1999 Ninth International Conference
  on Artificial Neural Networks ICANN 99. (Conf. Publ. No. 470)}, vol.~2,
  pp.~850--855 vol.2, Sep. 1999.

\bibitem{goodfellow2016}
I.~Goodfellow, Y.~Bengio, and A.~Courville, {\em Deep Learning}.
\newblock MIT Press, 2016.
\newblock \url{http://www.deeplearningbook.org}.

\bibitem{hinton2012neural}
G.~Hinton, N.~Srivastava, and K.~Swersky, ``Neural networks for machine
  learning lecture 6a overview of mini-batch gradient descent,'' {\em Cited
  on}, vol.~14, p.~8, 2012.

\bibitem{hastie2005elements}
T.~Hastie, R.~Tibshirani, J.~Friedman, and J.~Franklin, ``The elements of
  statistical learning: data mining, inference and prediction,'' {\em The
  Mathematical Intelligencer}, vol.~27, no.~2, pp.~83--85, 2005.

\bibitem{beckel2014eco}
C.~Beckel, W.~Kleiminger, R.~Cicchetti, T.~Staake, and S.~Santini, ``The eco
  data set and the performance of non-intrusive load monitoring algorithms,''
  in {\em Proceedings of the 1st ACM Conference on Embedded Systems for
  Energy-Efficient Buildings}, pp.~80--89, ACM, 2014.

\bibitem{mosley2013balanced}
L.~Mosley, {\em A balanced approach to the multi-class imbalance problem}.
\newblock PhD thesis, Iowa State University, 2013.

\bibitem{shen2019}
Y.~Shen, M.~Abubakar, H.~Liu, and F.~Hussain, ``Power quality disturbance
  monitoring and classification based on improved pca and convolution neural
  network for wind-grid distribution systems,'' {\em Energies}, vol.~12, no.~7,
  2019.

\bibitem{tieleman2012lecture}
T.~Tieleman and G.~Hinton, ``Lecture 6.5-rmsprop: Divide the gradient by a
  running average of its recent magnitude,'' {\em COURSERA: Neural networks for
  machine learning}, vol.~4, no.~2, pp.~26--31, 2012.

\bibitem{salehkalaibar2019}
S.~{Salehkalaibar}, F.~{Aminifar}, and M.~{Shahidehpour}, ``Hypothesis testing
  for privacy of smart meters with side information,'' {\em IEEE Transactions
  on Smart Grid}, vol.~10, no.~2, pp.~2059--2067, 2019.

\end{thebibliography}

\begin{IEEEbiography}[{\includegraphics[width=1in,height=1.25in,clip,keepaspectratio]{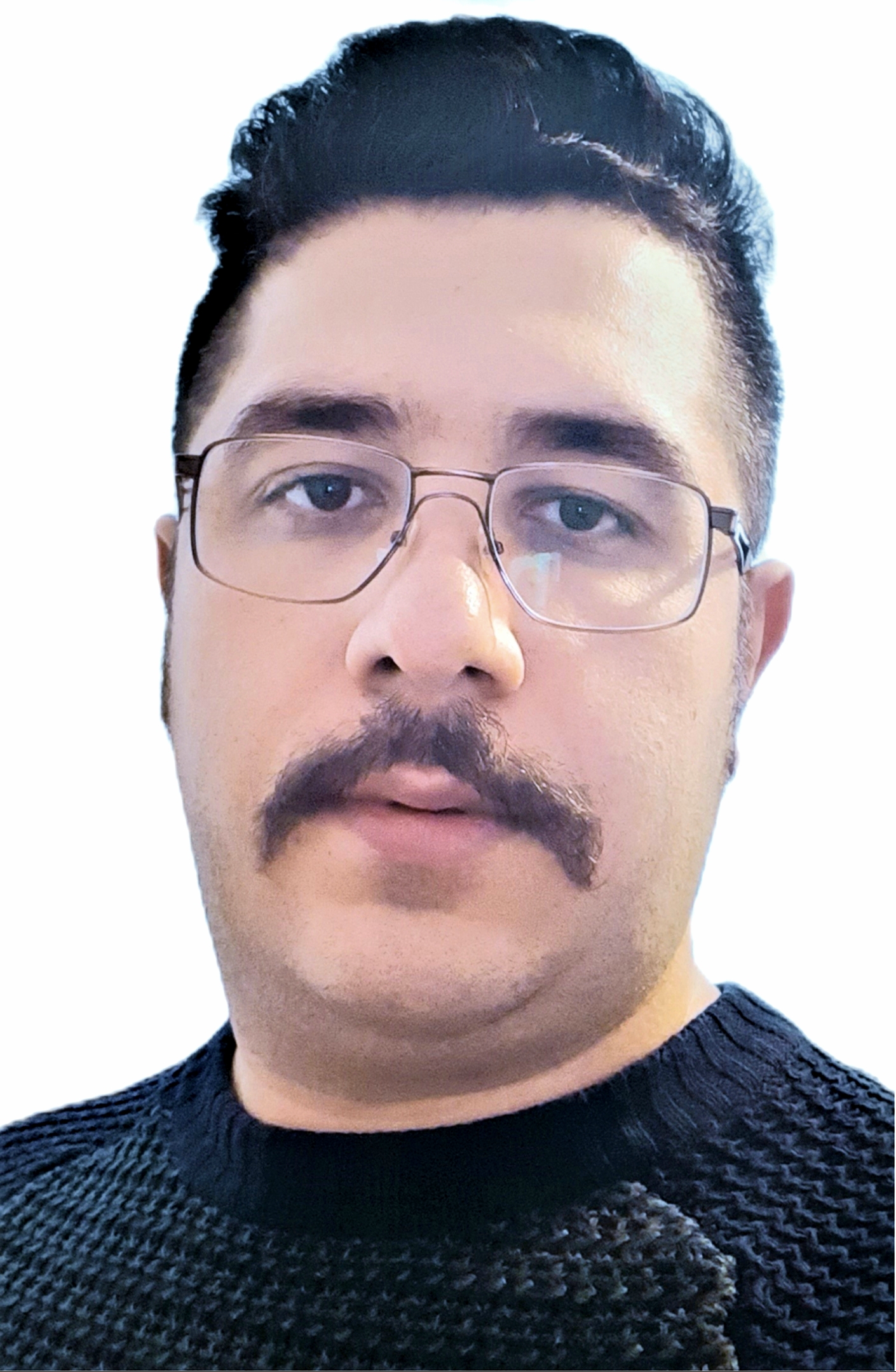}}]{Mohammadhadi Shateri}

(Member, IEEE) received the B.Sc. degree (with honors) in electrical engineering from the Amirkabir University of Technology, Tehran, Iran in 2012 and the M.Sc. degree (with honors) in electrical engineering from the University of Manitoba, Winnipeg, Canada in 2017. Currently, he is a Ph.D. candidate in electrical engineering at McGill University, Montreal, Canada. His research interests include machine learning, deep learning, and reinforcement learning with application to data analytics and smart grids.  

\end{IEEEbiography}

\begin{IEEEbiography}[{\includegraphics[width=1in,height=1.25in,clip,keepaspectratio]{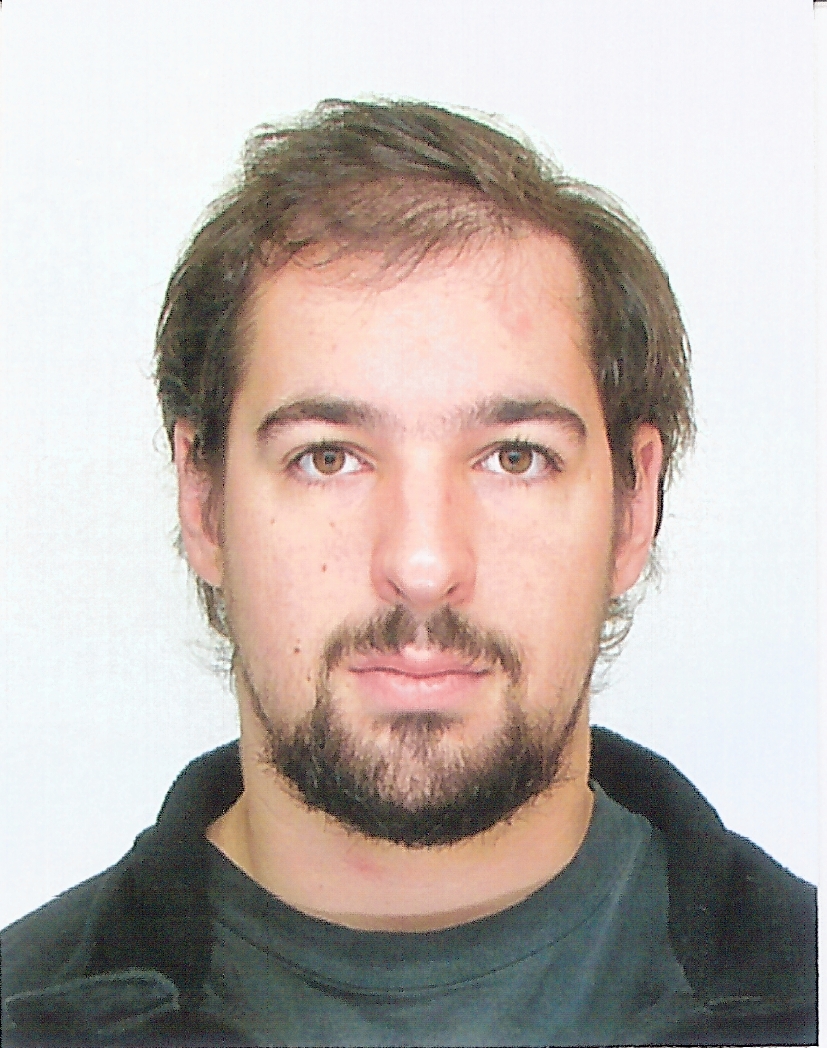}}]{Francisco Messina}

(Member, IEEE) (S’16, M’18) received the M.Sc. and Ph.D. (\emph{Summa Cum Laude}) degrees in electrical engineering from the University of Buenos Aires, Buenos Aires, Argentina, in 2014 and 2018, respectively. Currently, he is a Postdoctoral Fellow at McGill University, Montreal, Canada. His research interests include signal processing and machine learning with a focus on their applications to smart grids.

\end{IEEEbiography}

\begin{IEEEbiography}[{\includegraphics[width=1in,height=1.25in,clip,keepaspectratio]{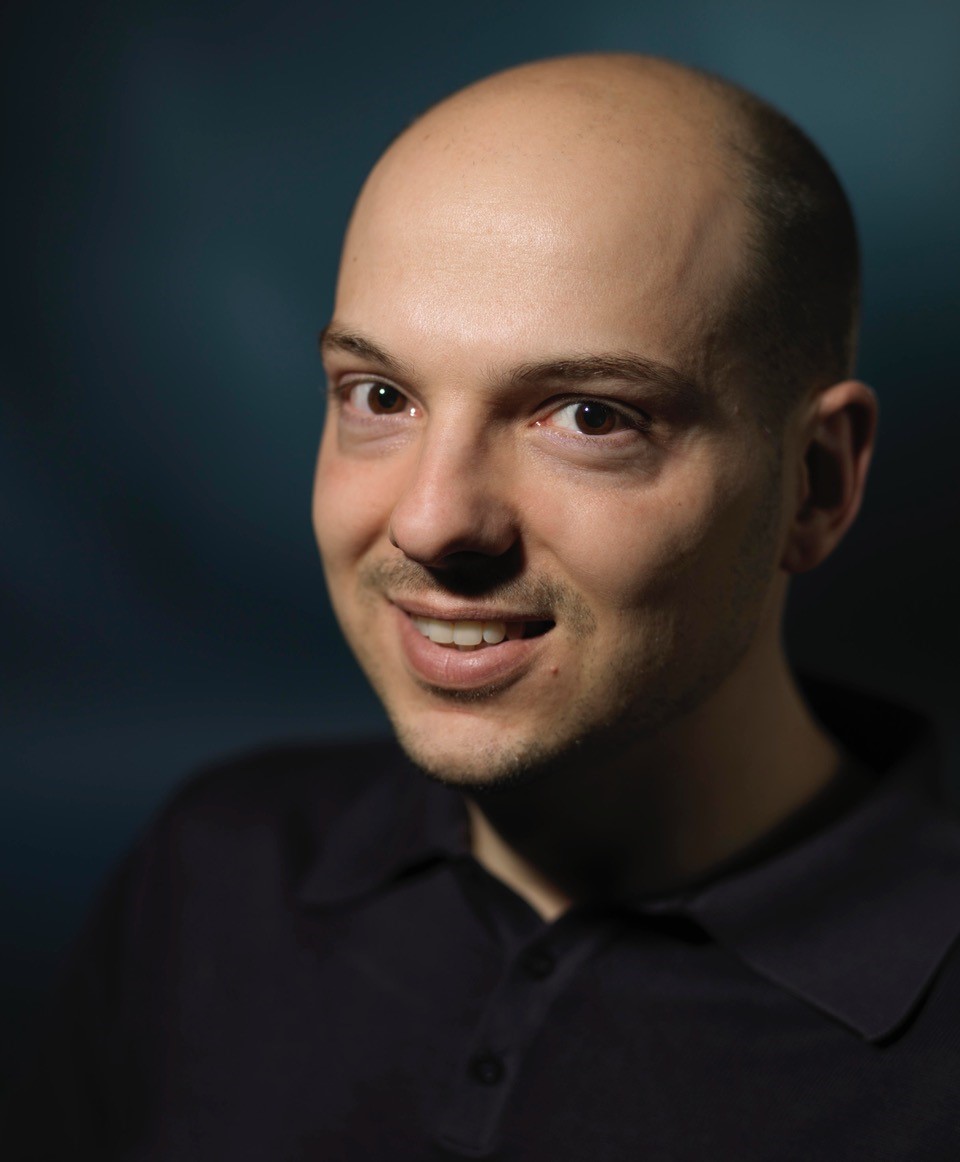}}]{Pablo Piantanida}

(Senior Member, IEEE) received the B.Sc. degree in electrical engineering and the M.Sc. degree from the University of Buenos Aires, Argentina, in 2003, and the Ph.D. degree from Universit\'e Paris-Sud, Orsay, France, in 2007. Since October 2007, he has been an Associate Professor of network information theory with the Laboratoire des Signaux et Syst\'emes (L2S), CentraleSup\'elec together with CNRS and Universit\'e Paris-Saclay. He is currently also with the Montreal Institute for Learning Algorithms (Mila), Universit\'e de Montr\'eal, QC, Canada. His research interests include information theory and its interactions with other fields, information measures, Shannon theory, machine learning and representation learning, statistical inference, and statistical  mechanisms for security and privacy. He has served as the General Co-Chair for the 2019 IEEE International Symposium on Information Theory (ISIT). He serves as an Associate Editor for the IEEE TRANSACTIONS ON INFORMATION FORENSICS AND SECURITY.

\end{IEEEbiography}

\begin{IEEEbiography}[{\includegraphics[width=1in,height=1.25in,clip,keepaspectratio]{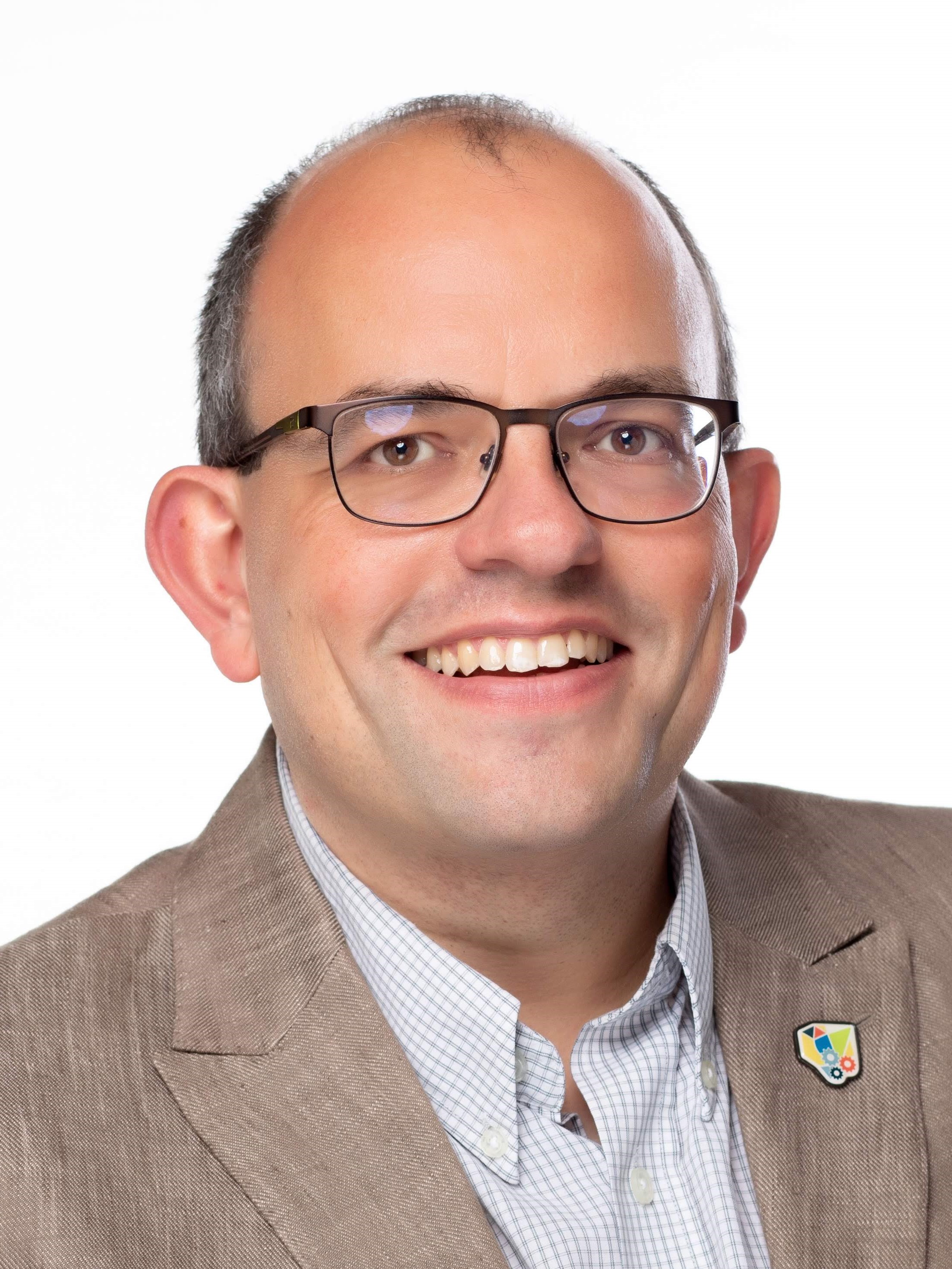}}]{Fabrice Labeau}

(Senior Member, IEEE) is the Deputy Provost (Student Life and Learning) at McGill University, where he also holds the NSERC/Hydro-Québec Industrial Research Chair in Interactive Information Infrastructure for the Power Grid. His research interests are in applications of signal processing. He has (co-)authored more than 200 papers in refereed journals and conference proceedings in these areas. He is the Director of Operations of STARaCom, an interuniversity research center grouping 50 professors and 500 researchers from 10 universities in the province of Quebec, Canada. He is Past President of the IEEE Sensors Council, a former President of the IEEE Vehicular Technology Society, ad former chair of the Montreal IEEE Section.  He was a recipient in 2015 and 2017 of the McGill University Equity and Community Building Award (team category), of the 2008 and 2016 Outstanding Service Award from the IEEE Vehicular Technology Society and of the 2017 W.S. Read Outstanding Service Award form IEEE Canada.

\end{IEEEbiography}

\end{document}